\documentclass[useAMS,usenatbib,usegraphicx]{mn2e}
\usepackage{amsmath}
\usepackage{color} 
\newcommand{\eqd}{\,\, .}
\newcommand{\eqc}{\,\, ,}
\newcommand{\pd}[1]{\, \partial #1 \,}
\newcommand{\td}[1]{\, \mathrm{d} #1 \,}
\newcommand{\intl}{\int\limits}

\newcommand{\HF}[1]{\; \mathrm{H}\left[ #1 \right]}  
\newcommand{\DF}[1]{\; \delta\left( #1 \right)}  
\newcommand{\eps}{\epsilon}
\newcommand{\lam}{\lambda}
\newcommand{\tl}{t_{lin}}
\newcommand{\latl}{\frac{\lam_0}{\tl}}
\newcommand{\tla}{\frac{\tl}{3\alpha^2}}
\newcommand{\latla}{\frac{3\alpha^2\lam_0}{\tl}}
\newcommand{\ks}{k_{s}}
\newcommand{\Gams}{\Gamma_{s}}
\newcommand{\kss}[1]{k_{s,#1}} 
\newcommand{\tss}[1]{t_{s,#1}} 
\newcommand{\Ls}[1]{L_{s}\left( #1 \right)}
\newcommand{\epss}{\epsilon_s}
\newcommand{\epec}{\epsilon_{ec}}
\newcommand{\Gamec}{\Gamma_{ec}}
\newcommand{\epsu}[1]{\eps_{e,#1}} 
\newcommand{\tsu}[1]{t_{e,#1}} 
\newcommand{\Le}{L_{e}\left( \epss,t \right)}
\newcommand{\sqrteeq}[1]{\sqrt{1+\frac{1}{\epss\epec q_{ #1 }}}}
\title[IC light curves of SSC cooled electrons]{Inverse Compton light curves of blazars under non-linear, time-dependent synchrotron-self Compton cooling}
\author[M. Zacharias]{M. Zacharias \\
Landessternwarte, Universit\"at Heidelberg, K\"onigstuhl, D-69117 Heidelberg, Germany \\
m.zacharias@lsw.uni-heidelberg.de}
\date{Received ?; accepted ? }
\begin{document}
\maketitle
\begin{abstract}
Blazars exhibit flares with a doubling time scale on the order of minutes. Such rapid flares are theoretically challenging and several {models} have been put forward to explain the fast variability. In this paper we continue the discussion concerning the effects of non-linear, time-dependent synchrotron self-Compton (SSC) cooling. In previous papers we were able to show that the non-linearity{, introduced by a time-dependent electron injection,} has severe consequences for both the spectral energy distribution (SED) and the monochromatic synchrotron light curve. The non-linear cooling introduces novel breaks in the SED, which are usually explained by complicated underlying electron distributions, while the much faster cooling of the SSC process {causes significant differences in the synchrotron light curves}. In this paper we calculate the inverse Compton light curves, taking into account both the SSC and the external Compton process. The light curves are calculated from the monochromatic intensities by introducing the retardation due to the finite size of the emission region and the geometry of the source. Even though some of the obvious effects of the SSC cooling are washed out by the retardation, there are still several observational constraints which could help to discriminate between the non-linear and the usual linear models, such as different flux states, temporal shapes or faster variability of the light curves at different energies.
\end{abstract}
%
%
\section{Introduction} \label{sec:intro}
Active galactic nuclei (AGN) belong to the most energetic classes of astrophysical objects, since they can be seen even in the far distances of the universe. {Especially blazars, a sub-class of AGN, characterize} these high energetic events. Blazars cover the entire electromagnetic emission spectrum from the very low radio frequencies up to very high energy $\gamma$-rays, divided into two broad non-thermal emission humps. According to the generally accepted AGN classification scheme (Urry \& Padovani, 1995), blazars are viewed at a very small angle with respect to the jet. The jet originates close to the central accreting super-massive black hole, and is a highly collimated, magnetized, relativistic {outflow} containing a mixture of electrons, positrons, and protons.

The standard {one-zone} model of blazar emission assumes a spherical, homogeneous emission region (also referred to as a blob) travelling at relativistic speeds down the jet. In the reference frame of the blob, the {relativistic particles contained in the emission region emit} powerful, non-thermal radiation. This emission is strongly boosted in the observer's frame due to the relativistic motion of the {blob} and the small viewing angle.

It is widely accepted that the low-energetic feature in a blazar spectrum is synchrotron emission from electrons (and probably positrons, which we imply from now on). In leptonic models (for reviews see B\"ottcher, 2007; 2012) the protons are assumed to be cold and {do not contribute} to the blazar emission (but see, e.g., B\"ottcher et al., 2013). The high energetic emission is then produced by inverse Compton scatterings of the electrons with ambient soft photon fields. Such photon fields can be either the internally self-produced synchrotron photons in the so-called synchrotron self-Compton (SSC) process (Jones et al., 1974), or the thermal photon fields, which are abundant external to the jet, in the so-called external Compton (EC) process. External photon fields can be the direct accretion disk radiation (Dermer \& Schlickeiser, 1993), the radiation of the broad line regions (Sikora et al., 1994), or the emission of the dusty torus (Blazejowski et al., 2000; Arbeiter et al., 2002). If the blazar emission occurs far away from the central engine, the cosmic microwave background could also serve as a soft target photon field {(B\"ottcher et al., 2008)}.

Blazars can exhibit very rapid flares with doubling time scales on the order of a few minutes as in the case of PKS 2155-304 (Aharonian et al., 2007{; 2009}), or PKS 1222+216 (Tavecchio et al., 2011) in the TeV range, or Mrk 421 (Cui, 2004) in the X-rays. Such short flares are theoretically challenging, since typical cooling and light crossing time scales are much longer. Several models have been invoked to explain these fast flares (e.g., Giannios et al., 2009; Ghisellini et al., 2009; Barkov et al., 2012; Biteau \& Giebels, 2012). Most of them have in common that they assume an emission region that is smaller than the cross-section of the jet, and moves much faster than the surrounding jet material. Thus, the smaller light crossing time scale in combination with stronger beaming effects explains the rapid flares.

However, the majority of these models also have in common that the underlying electron energy distribution function is stationary. Since particles are continuously injected {into} and lost from the source, the particle distribution will reach equilibrium, even if some parameters of the emission region change (in order to ``cause'' a flaring event). {On the other hand, several} investigations have shown that a flare caused by a {time-dependent} injection of particles into the emission region gives rise to substantially different effects, which decrease the cooling and variability time scales significantly, since the particles cannot reach equilibrium {(Li \& Kusunose, 2000; Katarzynski et al., 2006; Schlickeiser \& Lerche, 2007; Schlickeiser, 2009)}. 

Consider {a time-dependent scenario, where highly relativistic electrons are injected once into} an emission region pervaded by an isotropic magnetic field. The electrons inevitably produce synchrotron emission and {subsequently inverse Compton scatter these self-produced synchrotron photons (SSC process)}. Since no equilibrium can be established, the cooling of the electrons causes {a decrease in} the energy density in the synchrotron field, and thus the strength of the SSC process {decreases accordingly}. This implies a non-linear feedback, {since the SSC cooling term depends on the time-dependent} electron distribution function itself. Therefore, the SSC cooling efficiency decreases {with respect to time}. Schlickeiser (2009), and Zacharias \& Schlickeiser (2010) were able to show that this non-linear feedback substantially increases the electron cooling, implying much shorter variability time scales.

{The combination of the non-linear SSC cooling term with the linear, time-independent synchrotron and EC cooling terms has strong impacts on the electron distribution function. Due to the time-dependent decrease of the SSC cooling efficiency, the linear cooling terms will begin to dominate the cooling at a certain moment in time. Thus, the cooling behaviour of the source changes, which has strong consequences. For example, in the spectral energy distribution (SED) unique breaks emerge, which depend solely on the changing cooling behaviour (Schlickeiser et al., 2010; Zacharias \& Schlickeiser, 2012a; 2012b), and which would normally be} explained with complicated equilibrium electron distribution functions (in some cases without a thorough theoretical justification, e.g. Abdo et al., 2011). 

In a recent paper, Zacharias \& Schlickeiser (2013, hereafter referred to as ZS) concentrated on the effects of the combined linear and non-linear cooling on the emerging synchrotron light curves. They incorporated the retardation due to the finite size and the geometry of the source, which washed out some of the effects of the rapid non-linear cooling. Nevertheless, clear difference{s were shown} between the combined and the purely linear cooling. For instance, breaks in the rising part of the light curve before the light crossing time (LCT) occurred earlier in the combined cooling scenario than in the purely linear scenario. For variability taking place beyond the LCT at lower energies the respective variability time scales were also considerably shorter {in the combined scenario}. Thus, at least for the synchrotron emission, the non-linear, time-dependent cooling can potentially explain the rapid flares without the need for extreme assumptions.

In this paper we continue this analysis by calculating {analytically the} monochromatic light curves for the SSC and the EC emission. We will use the same set-up as ZS, which is outlined in section \ref{sec:app}. Section \ref{sec:int} presents the necessary parameters and the SSC and EC intensities. Details of the derivation of the intensities are given in appendix \ref{app:int}. In sections \ref{sec:ssca01} and \ref{sec:ssca10} the SSC light curves are calculated for purely linear and combined cooling scenarios, respectively, while the same is done for the EC light curves in sections \ref{sec:eca01} and \ref{sec:eca10}. We discuss the obtained analytical results and compare them to numerical integrations in section \ref{sec:dis}.

The purpose of this paper is to highlight the arising differences between the usual linear and the non-linear cooling scenarios. We therefore stick to a rather simple model, e.g. monochromatic injection of the electrons. The simplicity highlights the theoretical differences between the models, but the achieved results are probably too simple for a comparison with actual data. 
%
%
\section{The approach} \label{sec:app}
\begin{figure}
	\centering
		\includegraphics[width=0.48\textwidth]{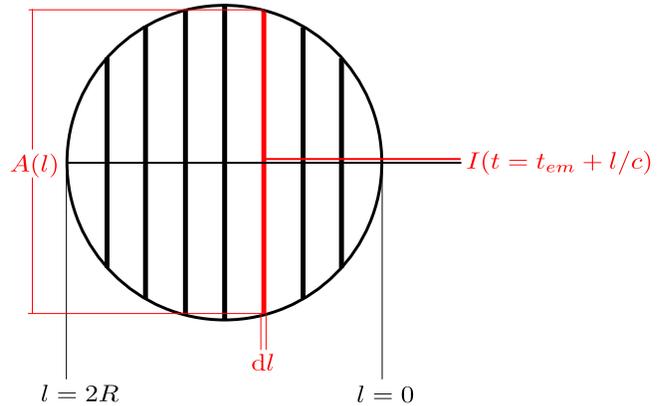}
	\caption{Sketch of the situation: The emission of the slice at position $l$ (with volume $\td{V(l)} = A(l)\td{l}$, where $A(l)$ is the position-dependent cross-section) is received by the observer {located to the right} at time $t=t_{em}+l/c$. Figure taken from ZS.}
	\label{fig:geo}
\end{figure}
We follow closely the steps of ZS (which are similar to, e.g. Chiaberge \& Ghisellini, 1999). The light curves are calculated from the respective intensities by incorporating the retardation due to the finite size of the source, as shown in Fig. \ref{fig:geo}. A photon emitted at the front of the source will arrive at the observer {at an earlier time} than a photon emitted at the back. Thus, even if the variability in a specific energy band is significantly shorter than the LCT scale, we expect photons to be detected until the LCT is reached, since photons at the back of the source take a longer time to reach the observer. Additionally, we take into account the geometry of the source, namely in our case the spherical geometry. Since in such a geometry every slice of the source has a different volume (c.f. Fig. \ref{fig:geo}), the amount of photons reaching the observer depends on the position of the specific emission slice in the source. This includes the assumption that the observer is located far away from the source implying all photons of a specific slice travel the same distance to the observer.{\footnote{Otherwise, the slices would have to be curved complicating the calculations by a lot.}

Since we use a spherical geometry, the actual position of a comoving observer is not important, since the emitted radiation is distributed isotropically. Due to the relativistic transformations, for a stationary observer the source becomes oblate and brighter depending on the viewing angle to the direction of motion of the emission blob. This, however, is incorporated by the beaming and Doppler corrections, which can be applied after the calculations. Since these effects do not alter the principle differences between the linear and non-linear scenario, they are of no concern for our discussion.\footnote{This might be different for other geometries. However, as discussed in ZS, our approach is strictly valid only for a spherical source.} }

Thus, according to ZS the monochromatic light curves are calculated with
\begin{align}
L(\epss,t) = 6\intl_0^1 I(\epss,t-\lam_0\lam) (\lam-\lam^2) \HF{t-\lam_0\lam} \td{\lam} \label{eq:lc0} \eqd
\end{align}
Here, $I(\epss,t_{em})$ is the monochromatic intensity at scattered energy $\epss$, which is normalized to the electron rest energy $m_ec^2$, and emission time $t_{em} = t-\lam_0\lam$. $t$ is the observer time located in front of the source (Fig. \ref{fig:geo}), and $\lam_0=2R/c$ is the LCT, with $R=10^{15}R_{15}$cm being the radius of the spherical source and $c$ the speed of light. The factor $6(\lam-\lam^2)$ is {the} geometrical weight function, introduced to take into account the different contributions of each slice due to its position $\lam$ in the source. {The integral with respect to $\lam$ over the geometrical weight function yields unity per definition.} 

{The Heaviside function in Eq. (\ref{eq:lc0}) is defined as $\HF{x}=1$ for $x>0$, and $\HF{x}=0$ otherwise. It implies that photons emitted by a specific slice at position $\lam$ can only be detected by the observer after they have crossed the distance from the position of the slice to the front of the source. Hence, only for $t>\lam_0$ all slices contribute to the light curve. The purpose of the Heaviside function is to make sure that especially in the beginning the retardation is correctly taken into account, since this is the main cause for the increase of the light curves at early times (c.f. Figs. \ref{fig:Lslog} - \ref{fig:Lelin} in section \ref{sec:dis}). 

By altering the Heaviside function, more realistic scenarios could be invoked. For example, by including a velocity-dependent function an internal shock scenario could be realised. However, such a scenario requires also an alteration of the kinetic equation of the electron distribution function (c.f. appendix \ref{app:int}) to include particle acceleration, diffusion and catastrophic losses. Such a thorough treatment can only be done numerically (e.g. Sokolov et al., 2004; Graff et al., 2008; B\"ottcher \& Dermer, 2010).}

{Since we intend to stick to an analytical derivation, the approach used in this paper is kept simple, and we have to leave aside potentially important processes. For example, we account for the retardation due to the finite size of the source, but we neglect the internal retardation of the SSC processes. If a synchrotron photon is scattered, it has travelled a short distance between creation and scattering.\footnote{{Under usual circumstances many synchrotron photons should be able to leave the source without scattering. In the extreme opposite case, where most of the synchrotron photons are scattered, the synchrotron component is exponentially reduced. This case, however, might not be tractable with the volume-averaged approach and numerical schemes involving the solution of a Fokker-Planck-like equation are necessary.}} This causes a delay in the onset of the SSC process (cooling) with respect to the synchrotron and external Compton processes (Chen et al., 2014).

Furthermore, the volume from which an electron (called the scattering electron in this paragraph) is supplied with synchrotron photons for SSC scattering increases with time until the LCT (Sokolov et al., 2004). Consider two synchrotron photons arriving at the same time at the scattering electron: The photon created at the edge of the volume has travelled a longer time (and, therefore, was created at an earlier time) than the photon created closer to the scattering electron. Thus, the photon from the edge of the volume stems from a higher energetic electron, since that electron did not have as much time to cool as the electron creating the photon closer to the scattering electron, as seen from the point of view of the scattering electron. Hence, the photon from the edge of the volume has a higher energy than the photon created nearby. Approximately, a higher incident photon energy implies a larger cross section of the IC scattering. Thus, the SSC process remains strong until the LCT (since until then photons with the highest possible energies are provided from the receding edge of the volume around the scattering electron) and does not decrease as much as if the internal retardation of the SSC process is neglected. The potential importance of this effect will be discussed by B\"ottcher et al. (in prep).} 

For the external Compton process such an intrinsic retardation is not expected, as long as the soft external photon source is stationary in time (Sokolov \& Marscher, 2005).

Further aspects are discussed in section 6.1 of ZS. To summarize, the acceleration time scale might have an impact, since it will contribute to the rising part of the light curve. It might cause an additional delay at higher energies, since particles have to reach their maximum energy causing the light curve for higher energies to be delayed with respect to lower energies. Such inclusions of acceleration, or the internal retardation, and other potentially important aspects, cannot be included in the analytical calculations of this paper for obvious reasons. They are incorporated in more and more realistic numerical schemes (e.g., Li \& Kusunose, 2000; Mimica et al., 2004; Weidinger \& Spanier, 2010; Joshi \& B\"ottcher, 2011; Chen et al., 2014).
%
%
\section{Monochromatic intensities} \label{sec:int}
In order to calculate the light curves of the SSC and EC processes with Eq. (\ref{eq:lc0}), we {require} the respective monochromatic intensities at emission time $t_{em}=t-\lam_0\lam$. From section \ref{sec:ssca01} on we will refer to the intensities as the unretarded light curves, since they would resemble the light curves for negligible retardation.

The detailed calculations {of the SSC and EC intensities} can be found in Zacharias \& Schlickeiser (2012a; 2012b), respectively. We present the essential steps in appendix \ref{app:int}.

Basically, the intensities of all radiation processes are divided into two cases by the injection parameter
\begin{align}
\alpha = \sqrt{\frac{A_0q_0}{D_0\left( 1+l_{ec} \right)}} \gamma_0  \label{eq:alpha} \eqd
\end{align}
Intensities for $\alpha\ll 1$ are influenced only by the linear and time-independent synchrotron and EC cooling, while intensities for $\alpha\gg 1$ are also influenced by the non-linear, time-dependent SSC cooling term. The latter causes the electrons to cool faster with respect to the linear cooling, and thus the variability might be significantly shortened. After the crossover time 
\begin{align}
t_c = \frac{\alpha^3-1}{3\alpha^2} \tl \label{eq:tc} \eqc
\end{align}
with the linear cooling time scale $\tl = \left[ D_0\left( 1+l_{ec} \right)\gamma_0 \right]^{-1}$, the SSC cooling strength has decreased {so much that the cooling becomes linear.}

In Eqs. (\ref{eq:alpha}) and (\ref{eq:tc}) the parameters are: $A_0 = 1.2\times 10^{-18} R_{15} b^2$cm$^3$s$^{-1}$, $D_0= 1.3\times 10^{-9} b^2$s$^{-1}$, and $q_0$ is the constant electron number density in the source. The electrons are injected simultaneously {at intrinsic time} $t_{em}=0$ homogeneously into the source with the monochromatic injection Lorentz factor $\gamma_0\gg 1$. The emission region is pervaded by a disordered magnetic field of strength $B=b$Gauss.

The parameter
\begin{align}
l_{ec} = \frac{4\Gamma_b^2}{3} \frac{u_{ec}^{\prime}}{u_B} \label{eq:lec} \eqc
\end{align}
with the bulk Lorentz factor $\Gamma_b$ of the emission region, and the energy densities of the soft external radiation source {$u_{ec}^{\prime}$ (angle-averaged in the blob frame)} and the magnetic field {$u_B$}, respectively. $l_{ec}$ marks the relative strength of the linear cooling terms. For $l_{ec}\gg 1$ the EC process dominates, while for $l_{ec}\ll 1$ the synchrotron process determines the linear cooling.

The soft external photon source is assumed to exhibit a line-like spectrum at monochromatic normalized energy $\epec$.

With these definitions the intensities of the respective processes can be derived.
\subsection{SSC intensities}
The SSC intensity has been calculated by Zacharias \& Schlickeiser (2012a). For $\alpha\ll 1$ it becomes
\begin{align}
I_{s}(\ks,t) =& I_{s,0} \ks^{1/3} \left( 1+\frac{t-\lam_0\lam}{t_{lin}} \right)^{4/3} e^{-\ks \left( 1+\frac{t-\lam_0\lam}{t_{lin}} \right)^{4}} \nonumber \\
&\times \HF{\frac{1}{\Gams\left(1+\frac{t-\lam_0\lam}{t_{lin}}\right)}-\ks} \label{eq:Issc0} \eqd
\end{align}
For $\alpha\gg 1$ the SSC intensity is
\begin{align}
I_{s}(\ks,t) =& I_{s,0} \ks^{1/3} \left( 1+3\alpha^2\frac{t-\lam_0\lam}{t_{lin}} \right)^{4/9} e^{-\ks \left( 1+3\alpha^2\frac{t-\lam_0\lam}{t_{lin}} \right)^{4/3}} \nonumber \\
&\times \HF{\frac{1}{\Gams\left(1+3\alpha^2\frac{t-\lam_0\lam}{t_{lin}}\right)^{1/3}}-\ks} \nonumber \\
&\times \HF{t_c-(t-\lam_0\lam)} \label{eq:Issc1} \eqc \\
I_{s}(\ks,t) =& I_{s,0} \ks^{1/3} \left( \alpha_g+\frac{t-\lam_0\lam}{t_{lin}} \right)^{4/3} e^{-\ks \left( \alpha_g+\frac{t-\lam_0\lam}{t_{lin}} \right)^{4}} \nonumber \\
&\times \HF{\frac{1}{\Gams\left(\alpha_g+\frac{t-\lam_0\lam}{t_{lin}}\right)}-\ks} \nonumber \\
&\times \HF{(t-\lam_0\lam)-t_c} \label{eq:Issc2} \eqd
\end{align}
Here we introduced the abbreviation $\ks=\epss/(\Gams\gamma_0)$, and $I_{s,0} = 189 R^2 P_0 \sigma_T a_0 q_0^2 m_ec^2 \Gams / (1600 \pi \gamma_0^3)$. $\Gams=4\eps_0\gamma_0$ is the SSC Klein-Nishina parameter, and $\alpha_g = (1+2\alpha^3)/3\alpha^2$. The other open parameters are given in appendix \ref{app:int}.
\subsection{EC intensities}
The EC intensity has been calculated by Zacharias \& Schlickeiser (2012b). It is for $\alpha\ll 1$
\begin{align}
I_{e}(\epss,t) =& I_{e,0} \epss \left( 1+\frac{t-\lam_0\lam}{\tl} \right)^2 G\left( q\left( \frac{t-\lam_0\lam}{\tl} \right) \right)  \nonumber \\
&\times \HF{\gamma_0-\gamma_{ec}\left( 1+\frac{t-\lam_0\lam}{\tl} \right)} \label{eq:Iec0} \eqc
\end{align}
while for $\alpha\gg 1$ one obtains
\begin{align}
I_{e}(\epss,t) =& I_{e,0} \epss \left( 1+3\alpha^2\frac{t-\lam_0\lam}{\tl} \right)^{2/3} G\left( q_1\left( 3\alpha^2\frac{t-\lam_0\lam}{\tl} \right) \right) \nonumber \\
&\times \HF{\gamma_0-\gamma_{ec}\left( 1+3\alpha^2\frac{t-\lam_0\lam}{\tl} \right)^{1/3}} \nonumber \\
&\times \HF{t_c-(t-\lam_0\lam)} \label{eq:Iec1} \eqc \\
I_{e}(\epss,t) =& I_{e,0} \epss \left( \alpha_g+\frac{t-\lam_0\lam}{\tl} \right)^2 G\left( q_2\left( \frac{t-\lam_0\lam}{\tl} \right) \right)  \nonumber \\
&\times \HF{\gamma_0-\gamma_{ec}\left( \alpha_g+\frac{t-\lam_0\lam}{\tl} \right)} \nonumber \\
&\times \HF{(t-\lam_0\lam)-t_c} \label{eq:Iec2} \eqd
\end{align}

Here $I_{e,0} = Rc\sigma_T q_0 u_{ec}^{\prime} \Gamma_b^2 / (4\pi \epec^2 \gamma_0^2)$, as well as
\begin{align}
q(\tau) =& \frac{\epss (1+\tau)^2}{\Gamec(\gamma_0-\epss (1+\tau))} \label{eq:q} \eqc \\
q_1(\tau) =& \frac{\epss (1+\tau)^{2/3}}{\Gamec(\gamma_0-\epss (1+\tau)^{1/3})} \label{eq:q1} \eqc \\
q_2(\tau) =& \frac{\epss (\alpha_g+\tau)^2}{\Gamec(\gamma_0-\epss (\alpha_g+\tau))} \label{eq:q2} \eqc 
\end{align}
the EC Klein-Nishina parameter
\begin{align}
\Gamec = 4\epec\gamma_0 \eqc
\end{align}
and
\begin{align}
\gamma_{ec} = \frac{\epss}{2} \left[ 1+\sqrt{1+\frac{1}{\epec\epss}} \right] \label{eq:gamec} \eqd
\end{align}
Further definitions, which are not essential for the following calculations, can also be found in appendix \ref{app:int}.
%
%
\section{SSC light curves for $\alpha\ll 1$} \label{sec:ssca01}
\subsection{Definitions} \label{sec:ssc01def}
Before we can integrate Eq. (\ref{eq:Issc0}) in Eq. (\ref{eq:lc0}), we must first discuss the Heaviside functions, since they give valuable information concerning the integration limits. The Heaviside function in Eq. (\ref{eq:Issc0}) can be transformed into
\begin{align}
\HF{\lam-\frac{t-\tss{00}}{\lam_0}} \eqc
\end{align}
serving as a lower limit for $t>\tss{00}$. Comparing it to the upper limit, it becomes larger than unity for $t>\lam_0+\tss{00}$, implying a cut off at $t=\lam_0+\tss{00}$. We defined
\begin{align}
\tss{00}(\ks) = \tl\left( \frac{1}{\Gams \ks}-1 \right) \label{eq:ts00} \eqd
\end{align}
Obviously, $\tss{00}<0$ for $\ks>\Gams^{-1}$. Since the latter is forbidden according to Eq. (\ref{eq:Issc0}), we find $\tss{00}>0$ for all energies $\ks$. Both $\lam_0$ and $\tss{00}$ mark a change in the integration limits of Eq. (\ref{eq:lc0}). Relating them, we find $\tss{00}<\lam_0$ for $\ks>\kss{00}$. Thus, the points in time when each integration limit is exchanged, depends on the value of $\ks$. The energy {$\kss{00}$ is defined as}
\begin{align}
\kss{00} = \frac{1}{\Gams\left( 1+\latl \right)} \label{eq:ks00} \eqd
\end{align}

Furthermore, we need to define two additional points in time, which are related to the unretarded light curve. The first one represents the time of the local maximum of the unretarded light curve:
\begin{align}
\tss{01}(\ks) = \tl\left[ \left( 3\ks \right)^{-1/4} -1 \right] \label{eq:ts01} \eqd
\end{align}
Hence, a local maximum is only attained for $\ks<1/3$. One can easily verify that $\tss{01}<\tss{00}$ for all energies. For $\tss{01}<\lam_0$, the energies must fulfil $\ks>\kss{01}$, with the latter being
\begin{align}
\kss{01} = \frac{1}{3\left( 1+\latl \right)^4} \label{eq:ks01} \eqd
\end{align}
The second and, as we will see, more important point in time is related to the argument of the exponential function of the unretarded light curve. Rewriting the argument as
\begin{align}
\ks \left( 1+\frac{t}{\tl} \right)^4 =& \ks+\ks \left[ \left( 1+\frac{t}{\tl} \right)^4 -1  \right] \nonumber \\
=& \ks+A(\ks,t) \label{eq:forts02} \eqc
\end{align}
the exponential starts to cut off the unretarded light curve once $A(\ks,t)>1$. Solving for $t$, we obtain the requirement $t>\tss{02}$, with
\begin{align}
\tss{02}(\ks) = \tl \left[ \left( 1+\ks^{-1} \right)^{1/4} -1 \right] \label{eq:ts02} \eqc
\end{align}
which is valid for all energies $\ks$. Relating this to $\tss{01}$, we find $\tss{02}>\tss{01}$ for all energies, which is, in fact, an expected result, since the cut-off should come after the local maximum. The relation $\tss{02}>\lam_0$ can only be fulfilled if $\ks<\kss{02}$, with
\begin{align}
\kss{02} = \frac{1}{\left( 1+\latl \right)^4 -1} \label{eq:ks02} \eqd
\end{align} 
The relation $\tss{02}>\tss{00}$ requires $\ks>\kss{03}$, where $\kss{03}$ is the real, positive solution of the equation
\begin{align}
\ks^4+\ks^3-\Gams^{-1} = 0 \label{eq:defks03} \eqd
\end{align}
The solution to this equation is difficult to achieve. For plotting purposes in section \ref{sec:dis} the value is derived numerically.

As a matter of fact, $\tss{02}$ is important for the following calculations of the light curve, since for $t>\tss{02}$ the slices at the front of source feel the exponential cut-off. This figurative speech implies merely a reduction of photons from the front slices, while the back slices still supply photons. Therefore, one can expect a break in the light curve. This is important only for $t<\lam_0$, since otherwise the time scales are much longer than the retardation, and its effects become negligible.

Thus, there are four principle calculations to be done for each case. Depending on the value of $\ks$, these results must be stitched afterwards giving the light curves for the respective energies.
\subsection{Calculations for $\ks<\kss{00}$}
In this case $\tss{00}>\lam_0$. For $t<\mathrm{min}(\tss{02},\lam_0)$ we can approximate $t-\lam_0\lam\ll \tl$, and the integral becomes
\begin{align}
\Ls{\ks,t} =& 6 I_{s,0} \ks^{1/3} \intl_0^{t/\lam0} \left( 1+\frac{t-\lam_0\lam}{t_{lin}} \right)^{4/3} e^{-\ks \left( 1+\frac{t-\lam_0\lam}{t_{lin}} \right)^{4}} \nonumber \\
& \times \left( \lam-\lam^2 \right) \td{\lam} \nonumber \\
\approx & 6 I_{s,0} \ks^{1/3} e^{-\ks} \intl_0^{t/\lam_0} \left( \lam-\lam^2 \right) \td{\lam} \nonumber \\
=& 3 I_{s,0} \ks^{1/3} e^{-\ks} \left( \frac{t}{\lam_0} \right)^2 \left[ 1-\frac{2t}{3\lam_0} \right] \label{eq:Ls011} \eqd
\end{align}

For the possible case $\tss{02}<t<\lam_0$, the approximation used above cannot be utilized. The detailed calculation is presented in appendix \ref{app:ssca01}. The result is approximately
\begin{align}
\Ls{\ks,t} \approx \frac{3 I_{s,0}\tl}{2\lam_0} \ks^{-2/3} e^{-\ks} \left( \frac{t}{\lam_0} \right) \left[ 1-\frac{t}{\lam_0} \right] \label{eq:Ls012} \eqd
\end{align}
As discussed above, for $t>\tss{02}$ the light curve exhibits a break by unity to a softer increase. Due to the mentioned stitching, the explicit form of this solution is not so important, apart from the linear time dependence.

For $\lam_0<t<\tss{00}$ we can approximate $t\gg\lam_0\lam$, which is the formal way of saying that the retardation is unimportant here. Then,
\begin{align}
\Ls{\ks,t} =& 6 I_{s,0} \ks^{1/3} \intl_0^1 \left( 1+\frac{t-\lam_0\lam}{t_{lin}} \right)^{4/3} e^{-\ks \left( 1+\frac{t-\lam_0\lam}{t_{lin}} \right)^{4}} \nonumber \\
& \times \left( \lam-\lam^2 \right) \td{\lam} \nonumber \\
\approx & I_{s,0} \ks^{1/3} \left( 1+\frac{t}{t_{lin}} \right)^{4/3} e^{-\ks \left( 1+\frac{t}{t_{lin}} \right)^{4}} \label{eq:Ls013} \eqd
\end{align}

The last case $\tss{00}<t<\tss{00}+\lam_0$ can be similarly approximated, giving
\begin{align}
\Ls{\ks,t} =& 6 I_{s,0} \ks^{1/3} \intl_{\frac{t-\tss{00}}{\lam_0}}^1 \left( 1+\frac{t-\lam_0\lam}{t_{lin}} \right)^{4/3} e^{-\ks \left( 1+\frac{t-\lam_0\lam}{t_{lin}} \right)^{4}} \nonumber \\
& \times \left( \lam-\lam^2 \right) \td{\lam} \nonumber \\
\approx & I_{s,0} \ks^{1/3} \left( 1+\frac{t}{t_{lin}} \right)^{4/3} e^{-\ks \left( 1+\frac{t}{t_{lin}} \right)^{4}} \nonumber \\
& \times \left[ 1-\left( \frac{t}{\tss{00}+\lam_0} \right)^2 \right] \label{eq:Ls014} \eqd
\end{align}
As stated above, the light curve is cut off for $t\geq\tss{00}+\lam_0$.
\subsection{Results for $\ks<\kss{00}$}
Here we stitch the results obtained in the last section, giving the light curves for all times.

For $\ks<\kss{01}<1/3$ we find
\begin{align}
\Ls{\ks,t} =& 3 I_{s,0} \ks^{1/3} \frac{\left( \frac{t}{\lam_0} \right)^2}{1+3\left( \frac{t}{\lam_0} \right)^2} \left( 1+\frac{t}{\tl} \right)^{4/3} \nonumber \\
& \times e^{-\ks\left( 1+\frac{t}{\tl} \right)^4} \left[ 1-\left( \frac{t}{\tss{00}+\lam_0} \right)^2 \right] \label{eq:Ls01ai} \eqc
\end{align}
and for $\kss{01}<\ks<\kss{02}<1/3$
\begin{align}
\Ls{\ks,t} =& 3 I_{s,0} \ks^{1/3} \frac{\left( \frac{t}{\lam_0} \right)^2}{\left( 1+\frac{t}{\tss{02}} \right)^{7/3}} \left( 1+\frac{t}{\tl} \right)^{4/3} \nonumber \\
& \times e^{-\ks\left( 1+\frac{t}{\tl} \right)^4} \left[ 1-\left( \frac{t}{\tss{00}+\lam_0} \right)^2 \right] \label{eq:Ls01aii} \eqc
\end{align}
while for $\kss{02}<\ks<1/3$
\begin{align}
\Ls{\ks,t} =& 3 I_{s,0} \ks^{1/3} e^{-\ks} \frac{\left( \frac{t}{\lam_0} \right)^2}{\left( 1+\frac{t}{\tss{02}} \right)^{7/3}} \left( 1+\frac{t}{\tl} \right)^{4/3} \nonumber \\
& \times \left[ 1-\frac{t}{\lam_0} \right] \label{eq:Ls01aiii} \eqd
\end{align}
Since for $\ks>\kss{02}$ the unretarded light curve cuts off long before the LCT, the retarded light curve (\ref{eq:Ls01aiii}) cuts off at $t=\lam_0$, since then even the slices at the back of the source are rapidly fading. This is different for the light curve (\ref{eq:Ls01aii}), since $\tss{02}$ is on the order of $\lam_0$. There the break at $\tss{02}$ is countered by the influence of the behaviour of the unretarded lightcurve around its local maximum. The light curve (\ref{eq:Ls01ai}) is obviously only influenced by the retardation until the LCT $\lam_0$, but afterwards the retarded and unretarded light curve are indistinguishable.

For $1/3<\ks<\kss{02}$ the light curve equals
\begin{align}
\Ls{\ks,t} =& 3 I_{s,0} \ks^{1/3} \frac{\left( \frac{t}{\lam_0} \right)^2}{1+3\left( \frac{t}{\lam_0} \right)^2} \left( 1+\frac{t}{\tl} \right)^{4/3} \nonumber \\
& \times e^{-\ks\left( 1+\frac{t}{\tl} \right)^4} \left[ 1-\left( \frac{t}{\tss{00}+\lam_0} \right)^2 \right] \label{eq:Ls01bi} \eqc
\end{align}
while for $\kss{02}<\ks$ one obtains
\begin{align}
\Ls{\ks,t} =& 3 I_{s,0} \ks^{1/3} e^{-\ks} \frac{\left( \frac{t}{\lam_0} \right)^2}{1+\frac{t}{\tss{02}}} \left[ 1-\frac{t}{\lam_0} \right] \label{eq:Ls01bii} \eqd
\end{align}
Here we have the at first sight unusual result that Eq. (\ref{eq:Ls01bi}) equals Eq. (\ref{eq:Ls01ai}) instead of Eq. (\ref{eq:Ls01aii}). However, the involved kinematics $\kss{02}>1/3$ imply $\lam_0\la 0.4\tl$, which means a very small emission region. Hence, the retardation effects become less important in this case.
\subsection{Calculations for $\ks>\kss{00}$}
Obviously, the calculations are quite similar to the previous case, and some results can be used again. The energy restriction implies $\tss{00}<\lam_0$. Hence, there is a possibility of a further break at $t=\tss{00}$, as well as at $t=\tss{02}$.

For $t<\mathrm{min}(\tss{00},\tss{02})$, the result of Eq. (\ref{eq:Ls011}) is valid here as well, giving
\begin{align}
\Ls{\ks,t} =& 3 I_{s,0} \ks^{1/3} e^{-\ks} \left( \frac{t}{\lam_0} \right)^2 \left[ 1-\frac{2t}{3\lam_0} \right] \label{eq:Ls021} \eqd
\end{align}

For $\tss{02}<t<\lam_0$ Eq. (\ref{eq:Ls012}) is a good approximation, yielding
\begin{align}
\Ls{\ks,t} \approx \frac{3 I_{s,0}\tl}{2\lam_0} \ks^{-2/3} e^{-\ks} \left( \frac{t}{\lam_0} \right) \left[ 1-\frac{t}{\lam_0} \right] \label{eq:Ls022} \eqd
\end{align}

For $\tss{00}<t<\lam_0$ we can use the approximation leading to Eq. (\ref{eq:Ls021}). With the different lower limit, the result becomes
\begin{align}
\Ls{\ks,t} =& 6 I_{s,0} \ks^{1/3} \intl_{\frac{t-\tss{00}}{\lam_0}}^{t/\lam0} \left( 1+\frac{t-\lam_0\lam}{t_{lin}} \right)^{4/3} e^{-\ks \left( 1+\frac{t-\lam_0\lam}{t_{lin}} \right)^{4}} \nonumber \\
& \times \left( \lam-\lam^2 \right) \td{\lam} \nonumber \\
\approx & 6 I_{s,0} \ks^{1/3} e^{-\ks} \frac{\tss{00}t}{\lam_0^2} \left[ 1-\frac{t}{\lam_0} \right] \label{eq:Ls023} \eqd
\end{align}
At $t=\tss{00}$ there is also a break by unity, similarly to the break at $t=\tss{02}$. Hence, the relation $\tss{02}\grole\tss{00}$ decides only at which point in time the break occurs. The specific form of Eqs. (\ref{eq:Ls022}) and (\ref{eq:Ls023}) is, again, not important due to the later stitching of the solutions.

For $\lam_0<t<\tss{00}+\lam_0$ one can use Eq. (\ref{eq:Ls014}), resulting in
\begin{align}
\Ls{\ks,t} =& I_{s,0} \ks^{1/3} \left( 1+\frac{t}{t_{lin}} \right)^{4/3} e^{-\ks \left( 1+\frac{t}{t_{lin}} \right)^{4}} \nonumber \\
& \times \left[ 1-\left( \frac{t}{\tss{00}+\lam_0} \right)^2 \right] \label{eq:Ls024} \eqd
\end{align}
\subsection{Results for $\ks>\kss{00}$}
Here the stitched solutions are separated by several energy conditions, First of all for $\ks<\kss{03}$ and $\ks>(2\Gams)^{-1}$, one obtains
\begin{align}
\Ls{\ks,t} =& 3 I_{s,0} \ks^{1/3} e^{-\ks} \frac{\left( \frac{t}{\lam_0} \right)^2}{1+\frac{t}{2\tss{00}}} \left[ 1-\frac{t}{\lam_0} \right] \label{eq:Ls01ci} \eqc
\end{align}
while for $\ks<\kss{03}$ and $\ks<(2\Gams)^{-1}$ the solution equals
\begin{align}
\Ls{\ks,t} =& 3 I_{s,0} \ks^{1/3} e^{-\ks} \frac{\left( \frac{t}{\lam_0} \right)^2 \left( 1+\frac{t}{\tl} \right)^{4/3}}{\left( 1+\frac{t}{\tss{02}} \right)^{7/3}} \left[ 1-\frac{t}{\lam_0} \right] \label{eq:Ls01cii} \eqd
\end{align}

On the other hand, for $\ks>\kss{03}$ and $\ks>\kss{02}$, the stitching yields
\begin{align}
\Ls{\ks,t} =& 3 I_{s,0} \ks^{1/3} e^{-\ks} \frac{\left( \frac{t}{\lam_0} \right)^2}{1+\frac{t}{\tss{00}}} \left[ 1-\frac{t}{\lam_0} \right] \label{eq:Ls01di} \eqc
\end{align}
whereas for $\ks>\kss{03}$ and $\ks<\kss{02}$
\begin{align}
\Ls{\ks,t} =& 3 I_{s,0} \ks^{1/3} \frac{\left( \frac{t}{\lam_0} \right)^2}{\left( 1+\frac{t}{\tss{00}} \right)^{7/3}} \left( 1+\frac{t}{\tl} \right)^{4/3} \nonumber \\
& \times e^{-\ks\left( 1+\frac{t}{\tl} \right)^4} \left[ 1-\left( \frac{t}{\tss{00}+\lam_0} \right)^2 \right] \label{eq:Ls01dii} \eqd
\end{align}

For most cases in the high energy domain $\ks>\kss{00}$, the light curve is cut off at $t=\lam_0$, which is not surprising. All solutions contain the break either at $\tss{00}$ or $\tss{02}$.
%
%
\section{SSC light curves for $\alpha\gg 1$} \label{sec:ssca10}
In this section we integrate the intensities (\ref{eq:Issc1}) and (\ref{eq:Issc2}), which are divided at the emission time $t_{em}=t_c$. Due to the retardation, there will be a time, when both early and late time solutions contribute to the final light curve. Thus, we delay the stitching of the individual solutions until the end of this section, and calculate first all possible cases.
\subsection{Early time limit}
\subsubsection{Definitions} \label{sec:ssca10etdef}
The Heaviside functions in Eq. (\ref{eq:Issc1}) both give a lower limit for the integral. The first one can be rewritten as
\begin{align}
\HF{\lam-\frac{t-\tss{10}}{\lam_0}} \eqc
\end{align}
with
\begin{align}
\tss{10}(\ks) = \tla \left[ \left( \Gams\ks \right)^{-3} -1 \right] \label{eq:ts10} \eqd
\end{align}
The second Heaviside function becomes
\begin{align}
\HF{\lam-\frac{t-t_c}{\lam_0}} \eqd
\end{align}
Obviously, the light curves in this case terminate either at $t=\tss{10}+\lam_0$ or at $t=t_c+\lam_0$, whichever of $\tss{10}$ or $t_c$ is smaller. Relating them as $t_c<\tss{10}$ gives the energy condition $\ks<\kss{c10}$, with
\begin{align}
\kss{c10} = \frac{1}{\Gams\alpha} \label{eq:ksc10} \eqd
\end{align}
For $\ks>\kss{c10}$, implying $\tss{10}<t_c$, the entire light curve cuts off before $t_c$. Hence, the intensity (\ref{eq:Issc2}) is not needed for high energies.

For high energies $\ks>\kss{c10}$ one further obtains $\ks>\kss{10}$ for $\tss{10}<\lam_0$. It follows
\begin{align}
\kss{10} = \frac{1}{\Gams \left( 1+\latla \right)^{1/3}} \label{eq:ks10} \eqd
\end{align}

For low energies $\ks<\kss{c10}$, the condition $t_c<\lam_0$ does not yield an energy constraint. However, due to the change in the lower integration limit, a break can be expected in the light curve. Thus, there are four general cases to be considered, depending on $\kss{c10}$ and either $t_c$ or $\kss{10}$, respectively.

Additionally, we require the conditions from the unretarded light curve. The local maximum of the unretarded light curve is attained at
\begin{align}
\tss{11}(\ks) = \tla \left[ \left( 3\ks \right)^{-3/4} -1 \right] \label{eq:ts11} \eqc
\end{align}
while the exponential cut-off, following the definition of section \ref{sec:ssc01def}, becomes important at
\begin{align}
\tss{12}(\ks) = \tla \left[ \left( 1+\ks^{-1} \right)^{3/4} -1 \right] \label{eq:ts12} \eqd
\end{align}
Relating these time scales to the time scales above yields several more energy constraints. For $\tss{11}<\lam_0$ follows $\ks>\kss{11}$, with
\begin{align}
\kss{11} = \frac{1}{3\left( 1+\latla \right)^{4/3}} \label{eq:ks11} \eqc
\end{align} 
for $\tss{11}<t_c$ one finds $\ks>(3\alpha^4)^{-1}$, and $\tss{11}<\tss{10}$ is valid for all energies.

Similarly, for $\tss{12}<\lam_0$ one obtains $\ks>\kss{12}$, with
\begin{align}
\kss{12} = \frac{1}{\left( 1+\latla \right)^{4/3} -1} \label{eq:ks12} \eqc
\end{align}
for $\tss{12}<t_c$ the energy obeys $\ks>\kss{14}$, with
\begin{align}
\kss{14} = \frac{1}{\alpha^4 -1} \label{eq:ks14} \eqc
\end{align}
while the requirement $\tss{12}<\tss{10}$ results in the inequality
\begin{align}
\ks^4+\ks^3-\Gams^{-4} < 0 \label{eq:defks13} \eqc
\end{align}
which is the same condition as for $\ks{03}$ in Eq. (\ref{eq:defks03}). The real, positive solution, which can be obtained again numerically, is therefore $\ks=\kss{13}$.

With these definitions, we can construct the different parts of the light curve.
\subsubsection{Calculations for $\ks<\kss{c10}$ and $t_c<\lam_0$}
In this case $t_c<\tss{10}$. For $t<t_c$ we find for $\ks<\kss{14}$ (i.e., $\tss{12}>t_c$),
\begin{align}
\Ls{\ks,t} =& 6 I_{s,0} \ks^{1/3} \intl_0^{t/\lam0} \left( 1+3\alpha^2\frac{t-\lam_0\lam}{t_{lin}} \right)^{4/9} \nonumber \\
& \times e^{-\ks \left( 1+3\alpha^2\frac{t-\lam_0\lam}{t_{lin}} \right)^{4/3}} \left( \lam-\lam^2 \right) \td{\lam} \nonumber \\
\approx & 6 I_{s,0} \ks^{1/3} e^{-\ks} \intl_0^{t/\lam_0} \left( \lam-\lam^2 \right) \td{\lam} \nonumber \\
=& 3 I_{s,0} \ks^{1/3} e^{-\ks} \left( \frac{t}{\lam_0} \right)^2 \left[ 1-\frac{2t}{3\lam_0} \right] \label{eq:LsIA1ai} \eqd
\end{align}

This is also the solution in the case $\ks>\kss{14}$ for times $t<\tss{12}$. For times $\tss{12}<t<t_c$, the result is approximately
\begin{align}
\Ls{\ks,t} \approx \frac{3 I_{s,0}\tl}{2\alpha^2\lam_0} \ks^{-2/3} e^{-\ks} \left( \frac{t}{\lam_0} \right) \left[ 1-\frac{t}{\lam_0} \right] \label{eq:LsIA1aii} \eqd
\end{align}
The detailed calculation is presented in appendix \ref{app:ssca10a}. As expected, the light curve breaks by unity at $t=\tss{12}$, again due to the decline in photon emission in the front slices.

For $t_c<t<\lam_0$ and $\ks<\kss{12}<\kss{14}$, we find
\begin{align}
\Ls{\ks,t} =& 6 I_{s,0} \ks^{1/3} \intl_{\frac{t-t_c}{\lam_0}}^{t/\lam0} \left( 1+3\alpha^2\frac{t-\lam_0\lam}{t_{lin}} \right)^{4/9} \nonumber \\
& \times e^{-\ks \left( 1+3\alpha^2\frac{t-\lam_0\lam}{t_{lin}} \right)^{4/3}} \left( \lam-\lam^2 \right) \td{\lam} \nonumber \\
\approx & 3 I_{s,0} \ks^{1/3} e^{-\ks} \left[ \lam^2-\frac{2}{3}\lam^3 \right]_{\frac{t-t_c}{\lam_0}}^{t/\lam0} \nonumber \\
\approx & 6 I_{s,0} \ks^{1/3} e^{-\ks} \frac{t_c}{\lam_0} \left( \frac{t}{\lam_0} \right) \left[ 1-\frac{t}{\lam_0} \right] \label{eq:LsIA1bi} \eqd
\end{align}
At $t=t_c$ there is also a break by unity to a softer increase of the light curve. In this case the break is not due to a decline in photon emission of the front slices, but rather that the front slices change to the linear emission process, which is not treated here. 

Eq. (\ref{eq:LsIA1bi}) is also part of the solution for $\kss{12}<\ks<\kss{14}$, namely for the time range $t_c<t<\tss{12}$. For $\tss{12}<t<\lam_0$ one can use Eq. (\ref{eq:LsIA1aii}) again.

For $\ks>\kss{14}$ the time range is again $t_c<t<\lam_0$. However, here $\tss{12}<t_c$, and, thus, we must use Eq. (\ref{eq:LsIA1aii}) instead of Eq. (\ref{eq:LsIA1bi}).

For $\lam_0<t<\lam_0+t_c$ the integral is again easily solved, and becomes
\begin{align}
\Ls{\ks,t} =& 6 I_{s,0} \ks^{1/3} \intl_{\frac{t-t_c}{\lam_0}}^1 \left( 1+3\alpha^2\frac{t-\lam_0\lam}{t_{lin}} \right)^{4/9} \nonumber \\
& \times e^{-\ks \left( 1+3\alpha^2\frac{t-\lam_0\lam}{t_{lin}} \right)^{4/3}} \left( \lam-\lam^2 \right) \td{\lam} \nonumber \\
\approx & I_{s,0} \ks^{1/3} \left( 1+3\alpha^2\frac{t}{t_{lin}} \right)^{4/9} e^{-\ks \left( 1+3\alpha^2\frac{t}{t_{lin}} \right)^{4/3}} \nonumber \\
& \times \left[ 1-\left( \frac{t}{t_c+\lam_0} \right)^2 \right] \label{eq:LsIA1c} \eqd
\end{align}
\subsubsection{Calculations for $\ks<\kss{c10}$ and $t_c>\lam_0$}
For $\ks<\kss{12}$ (i.e., $\tss{12}>\lam_0$), the time range $t<\lam_0$ is not further divided, and we obtain using Eq. (\ref{eq:LsIA1ai})
\begin{align}
\Ls{\ks,t} = 3 I_{s,0} \ks^{1/3} e^{-\ks} \left( \frac{t}{\lam_0} \right)^2 \left[ 1-\frac{2t}{3\lam_0} \right] \label{eq:LsIA2ai} \eqd
\end{align}

For $\ks>\kss{12}$ this first time interval is divided at $t=\tss{12}$. Thus, for $t<\tss{12}$ it is the same as Eq. (\ref{eq:LsIA2ai}), while for $\tss{12}<t<\lam_0$ Eq. (\ref{eq:LsIA1aii}) can be used.

The next case is the time interval $\lam_0<t<t_c$. The integration is simple, yielding
\begin{align}
\Ls{\ks,t} =& 6 I_{s,0} \ks^{1/3} \intl_0^1 \left( 1+3\alpha^2\frac{t-\lam_0\lam}{t_{lin}} \right)^{4/9} \nonumber \\
& \times e^{-\ks \left( 1+3\alpha^2\frac{t-\lam_0\lam}{t_{lin}} \right)^{4/3}} \left( \lam-\lam^2 \right) \td{\lam} \nonumber \\
\approx & I_{s,0} \ks^{1/3} \left( 1+3\alpha^2\frac{t}{t_{lin}} \right)^{4/9} e^{-\ks \left( 1+3\alpha^2\frac{t}{t_{lin}} \right)^{4/3}} \label{eq:LsIA2b} \eqd
\end{align}

In the interval $t_c<t<t_c+\lam_0$ the light curve is the same as Eq. (\ref{eq:LsIA1c}).
\subsubsection{Calculations for $\ks>\kss{c10}$ and $\ks>\kss{10}$}
The advantage in this section is that we can basically use the previous results and only need to care about the various cases depending on energy and time. Of course, the cut-off $t_c+\lam_0$ must be replaced by $\tss{10+\lam_0}$ here.

For $t<\tss{10}$ and $\ks>\kss{13}$ (i.e., $\tss{12}>\tss{10}$), the solution (\ref{eq:LsIA1ai}) can be used, giving
\begin{align}
\Ls{\ks,t} = 3 I_{s,0} \ks^{1/3} e^{-\ks} \left( \frac{t}{\lam_0} \right)^2 \left[ 1-\frac{2t}{3\lam_0} \right] \label{eq:LsIB1ai} \eqd
\end{align}

For $\ks<\kss{13}$ and $\ks>\kss{12}$ Eq. (\ref{eq:LsIB1ai}) is also valid for $t<\tss{12}$, while for $\tss{12}<t<\tss{10}$ we can use Eq. (\ref{eq:LsIA1aii}) giving
\begin{align}
\Ls{\ks,t} \approx \frac{3 I_{s,0}\tl}{2\alpha^2\lam_0} \ks^{-2/3} e^{-\ks} \left( \frac{t}{\lam_0} \right) \left[ 1-\frac{t}{\lam_0} \right] \label{eq:LsIB1aii} \eqd
\end{align}

For the time interval $\tss{10}<t<\lam_0$ we find for $\ks>\kss{13}$ and $\ks<\kss{12}$ (implying the total time interval) 
\begin{align}
\Ls{\ks,t} \approx 6 I_{s,0} \ks^{1/3} e^{-\ks} \frac{t_c}{\lam_0} \left( \frac{t}{\lam_0} \right) \left[ 1-\frac{t}{\lam_0} \right] \label{eq:LsIB1bi} \eqc
\end{align}
which is the same as Eq. (\ref{eq:LsIA1bi}).

For $\ks>\kss{12}$ the solution is two-fold, consisting of Eq. (\ref{eq:LsIB1bi}) and Eq. (\ref{eq:LsIB1aii}) divided at $t=\tss{12}$.

For $\ks<\ks{13}$ (i.e., $\tss{12}<\tss{10}$) the time interval is covered entirely by Eq. (\ref{eq:LsIB1aii}).

The time regime $\lam_0<t<\lam_0+\tss{10}$
\begin{align}
\Ls{\ks,t} \approx & I_{s,0} \ks^{1/3} \left( 1+3\alpha^2\frac{t}{t_{lin}} \right)^{4/9} e^{-\ks \left( 1+3\alpha^2\frac{t}{t_{lin}} \right)^{4/3}} \nonumber \\
& \times \left[ 1-\left( \frac{t}{\tss{10}+\lam_0} \right)^2 \right] \label{eq:LsIB1c} \eqc
\end{align}
derived similarly as Eq. (\ref{eq:LsIA1c}).
\subsubsection{Calculations for $\ks>\kss{c10}$ and $\ks<\kss{10}$}
Since the energy constraints imply $\tss{10}>\lam_0$, the only division below $t=\lam_0$ can be obtained if $\tss{12}<\lam_0$. Thus, for $\ks>\kss{12}$ we can use for $t<\tss{12}$ the solution (\ref{eq:LsIA1ai}), giving
\begin{align}
\Ls{\ks,t} = 3 I_{s,0} \ks^{1/3} e^{-\ks} \left( \frac{t}{\lam_0} \right)^2 \left[ 1-\frac{2t}{3\lam_0} \right] \label{eq:LsIB2ai} \eqd
\end{align}

For $\tss{12}<t<\lam_0$, obviously
\begin{align}
\Ls{\ks,t} \approx \frac{3 I_{s,0}\tl}{2\alpha^2\lam_0} \ks^{-2/3} e^{-\ks} \left( \frac{t}{\lam_0} \right) \left[ 1-\frac{t}{\lam_0} \right] \label{eq:LsIB2aii} \eqc
\end{align}
can be used.

For $\ks<\kss{12}$, the entire time interval is governed by Eq. (\ref{eq:LsIB2ai}).

The second time domain $\lam_0<t<\tss{10}$ is easily calculated, yielding
\begin{align}
\Ls{\ks,t} \approx I_{s,0} \ks^{1/3} \left( 1+3\alpha^2\frac{t}{t_{lin}} \right)^{4/9} e^{-\ks \left( 1+3\alpha^2\frac{t}{t_{lin}} \right)^{4/3}} \label{eq:LsIB2b} \eqc
\end{align}
while the time interval $\tss{10}<t<\tss{10}+\lam_0$ becomes
\begin{align}
\Ls{\ks,t} \approx & I_{s,0} \ks^{1/3} \left( 1+3\alpha^2\frac{t}{t_{lin}} \right)^{4/9} e^{-\ks \left( 1+3\alpha^2\frac{t}{t_{lin}} \right)^{4/3}} \nonumber \\
& \times \left[ 1-\left( \frac{t}{\tss{10}+\lam_0} \right)^2 \right] \label{eq:LsIB2c} \eqd
\end{align}
\subsection{Late time limit} \label{sec:ssc10lt}
\subsubsection{Definitions} 
Here Eq. (\ref{eq:Issc2}) must be integrated to give the light curves in the late time regime $t>t_c$. Inspecting the Heaviside functions in Eq. (\ref{eq:Issc2}) we find that for $t_c<t<t_c+\lam_0$ the upper limit is given by $(t-t_c)/\lam_0$ and by unity for later times. Furthermore for times $t>\tss{20}$ the lower limit is $(t-\tss{20})/\lam_0$, where
\begin{align}
\tss{20}(\ks) = \tl \left[ \left( \Gams\ks \right)^{-1} -\alpha_g \right] \label{eq:ts20} \eqd
\end{align}
For $\tss{20}<t_c$ the lower limit can be larger than the upper limit, which is obviously forbidden. This relation implies $\ks>(\Gams\alpha)^{-1}=\kss{c10}$. Thus, as discussed in section \ref{sec:ssca10etdef}, for energies $\ks>\kss{c10}$ the late time intensity does not contribute to the light curve. Additionally, for times $t>\tss{20}+\lam_0$ the light curve is cut off for all energies.

A further consequence of the above mentioned integration limits is that $t=\lam_0$ is not as important as in the early time limit. The change in the upper limit (corresponding to the retardation being either important or unimportant) is delayed to $t=t_c+\lam_0$. This can be related to $\tss{20}$ by $t_c+\lam_0<\tss{20}$ resulting in $\ks<\kss{c20}$ with
\begin{align}
\kss{c20} = \frac{1}{\Gams\left( \alpha+\latl \right)} \label{eq:ksc20} \eqd
\end{align}
This energy divides the integration cases, since the sequence of changing the integration limits depends on the relation between $t_c+\lam_0$ and $\tss{20}$.

Furthermore, we define the points in time and the related energies with respect to the local maximum and the exponential cut-off of the unretarded light curve. The local maximum is attained at
\begin{align}
\tss{21}(\ks) = \tl \left[ \left( 3\ks \right)^{-1/4} - \alpha_g \right] \label{eq:ts21} \eqc
\end{align}
while the exponential cut-off becomes important for
\begin{align}
\tss{22}(\ks) = \tl \left[ \left( 1+\frac{1}{\ks} \right)^{1/4} - \alpha_g \right] \label{eq:ts22} \eqd
\end{align}

Relating $\tss{21}<t_c+\lam_0$, the energy fulfills $\ks>\kss{21}$ with
\begin{align}
\kss{21} = \frac{1}{3\left( \alpha+\latl \right)^4} \label{eq:ks21} \eqd
\end{align}

Similarly, for $\tss{22}<t_c+\lam_0$ one obtains $\ks>\kss{22}$, where
\begin{align}
\kss{22} = \frac{1}{\left( \alpha+\latl \right)^4 -1} \label{eq:ks22}
\end{align}

Since, as before, $\tss{22}$ is important concerning breaks in the resulting light curve, it must be related to further points in time. For $\tss{22}<\tss{20}$, one finds $\ks<\kss{23}$, where $\kss{23}$ is the positive, real solution of the equation
\begin{align}
\ks^4+\ks^3-\Gams^{-1} = 0 \eqc
\end{align}
which is the same as in the previous cases. Hence $\kss{23} = \kss{13} = \kss{03}$. 

Furthermore, for $\tss{22}<t_c$ the energy is related to $\ks>\kss{24}$, with
\begin{align}
\kss{24} = \frac{1}{\alpha^4 -1} \label{eq:ks24} \eqc
\end{align}
and, obviously, $\kss{24} = \kss{14}$.
\subsubsection{Calculations for $\ks>\kss{c20}$}
In this case $\tss{20}<t_c+\lam_0$. For times $t_c<t<\tss{20}$ several cases have to be considered. For $\ks>\kss{22}$, $\ks<\kss{23}$, and $\ks>\kss{24}$, which implies $\tss{22}<t_c$, we find
\begin{align}
\Ls{\ks,t} \approx \frac{3 I_{s,0} \tl}{2\alpha^{5/3}\lam_0} \ks^{2/3} e^{-\alpha^4\ks} \left( \frac{t-t_c}{\lam_0} \right) \left[ 1-\frac{t}{t_c+\lam_0} \right] \label{eq:LsII1aiAa} \eqd
\end{align}
The detailed calculations are presented in appendix \ref{app:ssca10b}.

For $\ks<\kss{24}$ we find $\tss{22}>t_c$ implying two solutions in the interval $t_c<t<\tss{20}$. The first one for $t<\tss{22}$ equals
\begin{align}
\Ls{\ks,t} =& 6 I_{s,0} \ks^{1/3} \intl_0^{\frac{t-t_c}{\lam_0}} \left( \alpha_g+\frac{t-\lam_0\lam}{t_{lin}} \right)^{4/3} \nonumber \\
& \times e^{-\ks \left( \alpha_g+\frac{t-\lam_0\lam}{t_{lin}} \right)^{4}} \left( \lam-\lam^2 \right) \td{\lam} \nonumber \\
\approx & 6 I_{s,0} \ks^{1/3} e^{-\ks} \intl_0^{\frac{t-t_c}{\lam_0}} \left( \lam-\lam^2 \right) \td{\lam} \nonumber \\
=& 3 I_{s,0} \ks^{1/3} e^{-\ks} \left( \frac{t-t_c}{\lam_0} \right)^2 \left[ 1-\frac{t}{t_c+\lam_0} \right] \label{eq:LsII1aiAb1} \eqd
\end{align}
For $t>\tss{22}$ we can use Eq. (\ref{eq:LsII1aiAa}).

For $\ks>\kss{23}$ (implying $\tss{22}>\tss{20}$, and, thus $\ks<\kss{24}$), solution (\ref{eq:LsII1aiAb1}) can be used for the whole time interval. The same is true for the case $\ks<\kss{22}$.

The time interval $\tss{20}<t<t_c+\lam_0$ can also be divided according to the energy requirements. For $\ks>\kss{22}$ and $\ks<\kss{23}$, implying $\tss{22}<\tss{20}$, one can use the result (\ref{eq:LsII1aiAa}). The case $\ks>\kss{23}$ is twofold. For $\tss{20}<t<\tss{22}$ one can approximate
\begin{align}
\Ls{\ks,t} =& 6 I_{s,0} \ks^{1/3} \intl_{\frac{t-\tss{20}}{\lam_0}}^{\frac{t-t_c}{\lam_0}} \left( \alpha_g+\frac{t-\lam_0\lam}{t_{lin}} \right)^{4/3} \nonumber \\
& \times e^{-\ks \left( \alpha_g+\frac{t-\lam_0\lam}{t_{lin}} \right)^{4}} \left( \lam-\lam^2 \right) \td{\lam} \nonumber \\
\approx & 6 I_{s,0} \ks^{1/3} e^{-\ks} \intl_{\frac{t-\tss{20}}{\lam_0}}^{\frac{t-t_c}{\lam_0}} \left( \lam-\lam^2 \right) \td{\lam} \nonumber \\
\approx & 3 I_{s,0} \ks^{1/3} e^{-\ks} \frac{\tss{20}-t_c}{\lam_0} \left( \frac{t-t_c}{\lam_0} \right) \left[ 1-\frac{t}{t_c+\lam_0} \right] \label{eq:LsII1biB1} \eqc
\end{align}
and one can see that the point in time $t=\tss{20}$ induces a break by unity in the light curve. For the interval $\tss{22}<t<t_c+\lam_0$ the solution (\ref{eq:LsII1aiAa}) is valid, again.

For $\ks<\kss{22}$, which implies $\tss{22}>t_c+\lam_0$, result (\ref{eq:LsII1biB1}) is valid for the entire time interval.

The time interval $t_c+\lam_0<t<\tss{20}+\lam_0$, where the retardation is unimportant, can be approximated as
\begin{align}
\Ls{\ks,t} =& 6 I_{s,0} \ks^{1/3} \intl_{\frac{t-\tss{20}}{\lam_0}}^{1} \left( \alpha_g+\frac{t-\lam_0\lam}{t_{lin}} \right)^{4/3} \nonumber \\
& \times e^{-\ks \left( \alpha_g+\frac{t-\lam_0\lam}{t_{lin}} \right)^{4}} \left( \lam-\lam^2 \right) \td{\lam} \nonumber \\
\approx & 6 I_{s,0} \ks^{1/3} \left( \alpha_g+\frac{t}{\tl} \right)^{4/3} e^{-\ks\left( \alpha_g+\frac{t}{\tl} \right)^4} \nonumber \\
& \times \intl_{\frac{t-\tss{20}}{\lam_0}}^{1} \left( \lam-\lam^2 \right) \td{\lam} \nonumber \\
\approx & I_{s,0} \ks^{1/3} \left( \alpha_g+\frac{t}{\tl} \right)^{4/3} e^{-\ks\left( \alpha_g+\frac{t}{\tl} \right)^4} \nonumber \\
& \times \left[ 1-\left( \frac{t}{\tss{20}+\lam_0} \right)^2 \right] \label{eq:LsII1c} \eqd
\end{align}
\subsubsection{Calculations for $\ks<\kss{c20}$}
Here $\tss{20}>t_c+\lam_0$, and, thus, only in the first time interval $t_c<t<t_c+\lam_0$ the retardation is important, and several cases have to be considered.

For $\ks>\kss{22}$ and $\ks>\kss{24}$, implying $\tss{22}<t_c$, the approximation (\ref{eq:LsII1aiAa}) can be used. 

For $\ks<\kss{24}$, the time interval is further divided. For $t_c<t<\tss{22}$ Eq. (\ref{eq:LsII1aiAb1}) is valid, while for $\tss{22}<t<t_c+\lam_0$ Eq. (\ref{eq:LsII1aiAa}) serves as a good solution.

The last case of this first time interval is for $\ks<\kss{22}$, where Eq. (\ref{eq:LsII1aiAb1}) covers the entire interval.

The light curve in the time interval $t_c+\lam_0<t<\tss{20}$ becomes
\begin{align}
\Ls{\ks,t} =& 6 I_{s,0} \ks^{1/3} \intl_{0}^{1} \left( \alpha_g+\frac{t-\lam_0\lam}{t_{lin}} \right)^{4/3} \nonumber \\
& \times e^{-\ks \left( \alpha_g+\frac{t-\lam_0\lam}{t_{lin}} \right)^{4}} \left( \lam-\lam^2 \right) \td{\lam} \nonumber \\
\approx & I_{s,0} \ks^{1/3} \left( \alpha_g+\frac{t}{\tl} \right)^{4/3} e^{-\ks\left( \alpha_g+\frac{t}{\tl} \right)^4} \label{eq:LsII2b} \eqc
\end{align}
using the known approximation steps, when the retardation is unimportant.

The time interval $\tss{20}<t<\tss{20}+\lam_0$ is well approximated by Eq. (\ref{eq:LsII1c}).
\subsection{Results for $\alpha\gg 1$}
In this section we collect and combine the results presented before to give the most compact form of the light curves possible. Interestingly, the given energy requirements imply that the late time limit is only strictly necessary for energies $\ks<\kss{21}$. For higher energies the early time results can be used for all times. We will give the results similar as before in a tabular form, listing all energy (and other) requirements and then the corresponding light curve.

For $\ks<\kss{c10}$, $t_c<\lam_0$, and $\ks<\kss{14}$:
\begin{align}
\Ls{\ks,t} =& 3\ks^{1/3} e^{-\ks} \frac{\left( \frac{t}{\lam_0} \right)^2 \left( 1+3\alpha^2 \frac{t}{\tl} \right)^{4/9}}{\left( 1+\frac{t}{2t_c} \right)^{13/9} \left( 1+2\frac{t}{\lam_0} \right)} \nonumber \\
& \times \left[ 1-\left( \frac{t}{\lam_0+t_c} \right)^2 \right] \label{eq:Ls10a} \eqd
\end{align}

For $\ks<\kss{c10}$, $t_c<\lam_0$, and $\ks>\kss{14}$:
\begin{align}
\Ls{\ks,t} =& 3\ks^{1/3} e^{-\ks} \frac{\left( \frac{t}{\lam_0} \right)^2 \left( 1+3\alpha^2 \frac{t}{\tl} \right)^{4/9}}{\left( 1+\frac{t}{2\tss{12}} \right)^{13/9} \left( 1+3\frac{t}{\lam_0} \right)} \nonumber \\
& \times  \left[ 1-\left( \frac{t}{\lam_0} \right)^2 \right] \label{eq:Ls10b} \eqd
\end{align}

For $\ks<\kss{c10}$, $t_c>\lam_0$, and $\ks<\kss{12}$:
\begin{align}
\Ls{\ks,t} =& 3\ks^{1/3} \frac{\left( \frac{t}{\lam_0} \right)^2}{\left( 1+3 \left( \frac{t}{\lam_0} \right)^2 \right)}  \left( 1+3\alpha^2 \frac{t}{\tl} \right)^{4/9} \nonumber \\
& \times e^{-\ks\left( 1+3\alpha^2\frac{t}{\tl} \right)^{4/3}} \left[ 1-\left( \frac{t}{\lam_0+\tss{20}} \right)^2 \right] \label{eq:Ls10c} \eqd
\end{align}

For $\ks<\kss{c10}$, $t_c>\lam_0$, and $\ks>\kss{12}$:
\begin{align}
\Ls{\ks,t} =& 3\ks^{1/3} e^{-\ks} \frac{\left( \frac{t}{\lam_0} \right)^2}{\left( 1+\frac{t}{\tss{12}} \right)} \left( 1+3\alpha^2 \frac{t}{\tl} \right)^{4/9} \nonumber \\
& \times \left[ 1-\frac{t}{\lam_0} \right] \label{eq:Ls10d} \eqd
\end{align}

For $\ks>\kss{c10}$, and $\ks>\kss{10}$:
\begin{align}
\Ls{\ks,t} =& 3\ks^{1/3} e^{-\ks} \frac{\left( \frac{t}{\lam_0} \right)^2 \left( 1+3\alpha^2 \frac{t}{\tl} \right)^{4/9}}{\left( 1+\frac{t}{\tss{12}} \right) \left( 1+3\frac{t}{\lam_0} \right)} \nonumber \\
& \times \left[ 1-\left( \frac{t}{\lam_0} \right)^2 \right] \label{eq:Ls10e} \eqd
\end{align}

For $\ks>\kss{c10}$, $\ks<\kss{10}$, and $\ks<\kss{12}$:
\begin{align}
\Ls{\ks,t} =& 3\ks^{1/3} \frac{\left( \frac{t}{\lam_0} \right)^2}{\left( 1+3 \left( \frac{t}{\lam_0} \right)^2 \right)}  \left( 1+3\alpha^2 \frac{t}{\tl} \right)^{4/9} \nonumber \\
& \times e^{-\ks\left( 1+3\alpha^2\frac{t}{\tl} \right)^{4/3}} \left[ 1-\left( \frac{t}{\lam_0+\tss{20}} \right)^2 \right] \label{eq:Ls10f} \eqd
\end{align}

For $\ks>\kss{c10}$, $\ks<\kss{10}$, and $\ks>\kss{12}$:
\begin{align}
\Ls{\ks,t} =& 3\ks^{1/3} e^{-\ks} \frac{\left( \frac{t}{\lam_0} \right)^2}{\left( 1+\frac{t}{2\tss{12}} \right)} \left[ 1-\left( \frac{t}{\lam_0} \right)^2 \right] \label{eq:Ls10g} \eqd
\end{align}

Finally, for $\ks<\kss{21}$, the only one where the late time limit is included:
\begin{align}
\Ls{\ks,t} =& 3\ks^{1/3} \frac{\left( \frac{t}{\lam_0} \right)^2}{\left( 1+3 \left( \frac{t}{\lam_0} \right)^2 \right)} \left[ 1-\left( \frac{t}{\lam_0+\tss{20}} \right)^2 \right] \nonumber \\
& \times \left\{ \left( 1+3\alpha^2 \frac{t}{\tl} \right)^{4/9} e^{-\ks\left( 1+3\alpha^2\frac{t}{\tl} \right)^{4/3}} \right. \nonumber \\
& \times \HF{t_c-t} \nonumber \\
& + \left( \alpha_g+\frac{t}{\tl} \right)^{4/3} e^{-\ks\left( \alpha_g+\frac{t}{\tl} \right)^{4}} \nonumber \\
& \times \left. \HF{t-t_c} \right\} \label{eq:Ls10h} \eqd
\end{align}

This stitching procedure has the advantage that the light curves are presented in a single equation. On the other hand, this causes the analytical results to be bad approximations in {some} cases, since not all details of the different energy cases could be covered. Additionally, the last light curve (\ref{eq:Ls10h}) could not be completely stitched, since otherwise the upturn for times $t>t_c$ could not be reproduced well. This is an artefact of the approximation used in the derivation of the electron distribution function (Schlickeiser et al., 2010), and even influences the numerical result, since the integrals containing Eqs. (\ref{eq:Issc1}) and (\ref{eq:Issc2}) are evaluated separately, of course. Nevertheless, the basic characteristics of the numerical result are achieved: Namely, for $t<\lam_0$ the retardation and geometry of the source dominate the light curve, while for later times the unretarded light curve is recovered. This will be discussed in detail in section \ref{sec:dis}.
%
%
\section{EC light curves for $\alpha\ll 1$} \label{sec:eca01}
\subsection{Definitions} \label{sec:ec01def}
The calculations of the EC light curve are similar to the SSC light curve, with the exception that we need fewer definitions and thresholds due to the simpler form of the intensity (\ref{eq:Iec0}). The Heaviside function in Eq. (\ref{eq:Iec0}) can be rewritten as
\begin{align}
\HF{\lam-\frac{t-\tsu{00}}{\lam_0}} \eqc
\end{align}
giving a lower limit for the integral in eq. (\ref{eq:lc0}) for $t>\tsu{00}$. The time
\begin{align}
\tsu{00} = \tl\left( \frac{\gamma_0}{\gamma_{ec}} - 1 \right) \label{eq:te00}
\end{align}
marks the cut-off of the unretarded light curve, beyond which no further photons are produced. The term in brackets implies an upper limit for the scattered photon energy $\epss<\Gamec\gamma_0/(1+\Gamec)$. 

The point in time $\tsu{00}$ can be related to $\lam_0$ and $\tl$ giving important energy constraints for the resulting light curve. For $\tsu{00}<\lam_0$, implying the cut-off of the unretarded light curve happening before the LCT is reached, the scattered energies must exceed $\epss>\epsu{00}$, where
\begin{align}
\epsu{00} = \frac{\Gamec\gamma_0}{\left( 1+\frac{\lam_0}{\tl} \right) \left( 1+\frac{\lam_0}{\tl} + \Gamec \right)} \label{eq:ee00} \eqd
\end{align}

Important only if $\tl<\lam_0$, the relation $\tsu{00}<\tl$ results in the energy relation $\epss>\epsu{01}$, with
\begin{align}
\epsu{01} = \frac{\Gamec\gamma_0}{4+2\Gamec} \label{eq:ee01} \eqd
\end{align}
For $\tl<\lam_0$ this energy is larger than $\epsu{00}$.
\subsection{Calculations for $\epss<\epsu{00}$}
In this case $\tsu{00}>\lam_0$ and similar approximations as in the calculation for the SSC light curves can be applied. For $t\ll \lam_0$ we approximate $t-\lam_0\lam\ll\tl$, giving
\begin{align}
\Le =& 6 I_{ec,0} \epss \intl_0^{t/\lam_0} \left( 1+\frac{t-\lam_0\lam}{\tl} \right)^2 G\left( q\left( \frac{t-\lam_0\lam}{\tl} \right) \right) \nonumber \\
&\times \left( \lam-\lam^2 \right) \td{\lam} \nonumber \\
\approx & 6 I_{ec,0} G\left( q(0) \right) \epss \intl_0^{t/\lam_0} \left( \lam-\lam^2 \right) \td{\lam} \nonumber \\
=& 3 I_{ec,0} G\left( q(0) \right) \epss \left( \frac{t}{\lam_0} \right)^2 \left[ 1-\frac{2t}{3\lam_0} \right] \label{eq:Le011} \eqd
\end{align}

For times $t\gg \lam_0$ the retardation is unimportant and we find for $\lam_0<t<\tsu{00}$
\begin{align}
\Le =& 6 I_{ec,0} \epss \intl_0^{1} \left( 1+\frac{t-\lam_0\lam}{\tl} \right)^2 G\left( q\left( \frac{t-\lam_0\lam}{\tl} \right) \right) \nonumber \\
&\times  \left( \lam-\lam^2 \right) \td{\lam} \nonumber \\
\approx & 6 I_{ec,0} \epss \left( 1+\frac{t}{\tl} \right)^2 G\left( q\left( \frac{t}{\tl} \right) \right) \intl_0^1 \left( \lam-\lam^2 \right) \td{\lam} \nonumber \\
=& I_{ec,0} \epss \left( 1+\frac{t}{\tl} \right)^2 G\left( q\left( \frac{t}{\tl} \right) \right) \label{eq:Le012} \eqc
\end{align}
while for $\tsu{00}<t<\tsu{00}+\lam_0$ the light curve becomes
\begin{align}
\Le =& 6 I_{ec,0} \epss \intl_{\frac{t-\tsu{00}}{\tl}}^{1} \left( 1+\frac{t-\lam_0\lam}{\tl} \right)^2 G\left( q\left( \frac{t-\lam_0\lam}{\tl} \right) \right) \nonumber \\
&\times  \left( \lam-\lam^2 \right) \td{\lam} \nonumber \\
\approx & 6 I_{ec,0} \epss \left( 1+\frac{t}{\tl} \right)^2 G\left( q\left( \frac{t}{\tl} \right) \right) \nonumber \\
&\times \intl_{\frac{t-\tsu{00}}{\tl}}^1 \left( \lam-\lam^2 \right) \td{\lam} \nonumber \\
\approx & I_{ec,0} \epss \left( 1+\frac{t}{\tl} \right)^2 G\left( q\left( \frac{t}{\tl} \right) \right) \nonumber \\
&\times \left[ 1-\frac{t}{\tsu{00}+\lam_0} \right] \label{eq:Le013} \eqd
\end{align}
\subsection{Calculations for $\epss>\epsu{00}$}
Here $\tsu{00}<\lam_0$ and we expect a break in the light curve at $t=\tsu{00}$, similar to the SSC cases. For times $t<\tsu{00}$ we can use Eq. (\ref{eq:Le011}). 

For times $\tsu{00}<t<\lam_0$ the detailed calculations are presented in appendix \ref{app:eca01}, and the solution equals
\begin{align}
\Le =& \frac{3}{8} I_{ec,0} \Gamec^3 \epss F_0(\epss) \frac{\tl}{\lam_0} \left( \frac{t}{\lam_0} \right) \left[ 1-\frac{t}{\lam_0} \right] \label{eq:Le014} \eqd
\end{align}
The explicit form of the integral $F_0(\eps)$ is not important, since this solution will be stitched to the previous solution. The linear time-dependence of the light curve in this time domain is the major result here, giving again a break by unity above the cut-off time.
\subsection{Results for $\alpha\ll 1$}
Combining the above results, and taking into account that for $\tl<\lam_0$ we have the additional energy requirement $\epsu{01}$, the resulting light curves for $\alpha\ll 1$ can be summarized as follows.

For $\epss<\epsu{00}$ we find
\begin{align}
\Le = 3 I_{ec,0} \epss \frac{\left( \frac{t}{\lam_0} \right)^2}{1+3\left( \frac{t}{\lam_0} \right)^2} \left( 1+\frac{t}{\tl} \right)^2 G\left( q\left( \frac{t}{\tl} \right) \right) \label{eq:Le01} \eqd
\end{align} 
For $\epsu{00}<\epss<\epsu{01}$ the light curve
\begin{align}
\Le = 3 I_{ec,0} \epss G\left( q(0) \right) \frac{\left( \frac{t}{\lam_0} \right)^2 \left( 1+\frac{t}{\tl} \right)^2}{\left( 1+\frac{t}{\tsu{00}} \right)^3} \left[ 1+ \left( \frac{t}{\lam_0} \right)^2 \right] \label{eq:Le02} \eqc
\end{align}
whereas for $\epss>\epsu{01}$
\begin{align}
\Le = 3 I_{ec,0} \epss G\left( q(0) \right) \frac{\left( \frac{t}{\lam_0} \right)^2}{1+\frac{t}{2\tsu{00}}} \left[ 1-\frac{t}{\tsu{00}+\lam_0} \right] \label{eq:Le03} \eqd
\end{align}

These solutions are {good} representations of the actual light curves for a large range of the parameter space (see section \ref{sec:dis}). Some cases are however not well covered with these approximations, since the breaks introduced by the stitching are less pronounced than in the numerical integration. Still, the basic characteristics of the light curves are represented, namely that for early times the retardation and the geometry determine the structure of the light curve, while beyond the LCT the retardation is almost unimportant.
%
%
\section{EC light curves for $\alpha\gg 1$} \label{sec:eca10}
In this section we calculate the EC light curves from Eqs. (\ref{eq:Iec1}) and (\ref{eq:Iec2}). First, we will calculate the light curve in the early time limit and then proceed to the late time limit. In the end, these solutions are stitched as much as possible to give single light curves for the entire time range.
\subsection{Early time limit}
\subsubsection{Definitions} \label{sec:ec10etdef}
As in the previous calculations at first some definitions are introduced, which are needed to differentiate between the various cases. From Eq. (\ref{eq:Iec1}) we find the cut-off time
\begin{align}
\tsu{10} = \frac{\tl}{3\alpha^2} \left[ \left( \frac{\gamma_0}{\gamma_{ec}} \right)^3 - 1 \right] \label{eq:te10}
\end{align}
by rearranging the second Heaviside function for $\lam$. Relating this to $t_c$, one finds for $\tsu{10}<t_c$
\begin{align}
\epss>\epsu{c10} = \frac{\Gamec\gamma_0}{\alpha\left( \alpha+\Gamec \right)} \label{eq:eec10} \eqd
\end{align}

For $\tsu{10}<\lam_0$ the energy condition becomes
\begin{align}
\epss>\epsu{10} = \frac{\Gamec\gamma_0}{\left( 1+3\alpha^2\frac{\lam_0}{\tl} \right)^{1/3} \left[ \left( 1+3\alpha^2\frac{\lam_0}{\tl} \right)^{1/3} + \Gamec \right]} \label{eq:ee10} \eqd
\end{align}
Lastly, $\tsu{10}<\tl/3\alpha^2$ yields
\begin{align}
\epss>\epsu{11} = \frac{\Gamec\gamma_0}{4^{1/3}+2^{1/3} \Gamec} \label{eq:ee11} \eqd
\end{align}
\subsubsection{Calcualtions for $\epss<\epsu{c10}$ and $t_c<\lam_0$}
Inserting Eq. (\ref{eq:Iec1}) in Eq. (\ref{eq:lc0}), we find for times $t<t_c$
\begin{align}
\Le =& 6 I_{e,0} \epss \intl_{0}^{t/\lam_0} \left( 1+3\alpha^2\frac{t-\lam_0\lam}{\tl} \right) \nonumber \\
&\times G\left( q_1\left( 3\alpha^2\frac{t-\lam_0\lam}{\tl} \right) \right) (\lam-\lam^2) \td{\lam} \nonumber \\
\approx & 6 I_{e,0} \epss G\left( q_1(0) \right) \intl_0^{t/\lam_0} (\lam-\lam^2) \td{\lam} \nonumber \\
=& 3 I_{e,0} \epss G\left( q_1(0) \right) \left( \frac{t}{\lam_0} \right)^2 \left[ 1-\frac{2t}{3\lam_0} \right] \label{eq:Le101} \eqd
\end{align}
The approximation is strictly valid only for $3\alpha^2(t-\lam_0\lam)\ll \tl$. This approximation cannot be used in the second time interval $t_c<t<\lam_0$. The detailed calculations are outlined in appendix \ref{app:eca10a} yielding
\begin{align}
\Le =& \frac{3I_{e,0} \Gamec^5}{32 \alpha^2} \epss F_1(\epss) \frac{\tl}{\lam_0} \left( \frac{t}{\lam_0} \right) \left[ 1-\frac{t}{\lam_0} \right] \label{eq:Le102} \eqd
\end{align}
As for Eq. (\ref{eq:Le014}) the explicit form of the integral $F_1(\epss)$ is unimportant for the present discussion. The time $t_c$ introduces a break, since the front slices of the emission region begin to cool linearly, which is calculated in section \ref{sec:ec10lt}.

For late times $\lam_0<t<\lam_0+t_c$ the retardation is approximately negligible. Hence,
\begin{align}
\Le =& 6 I_{e,0} \epss \intl_{\frac{t-t_c}{\lam_0}}^{1} \left( 1+3\alpha^2\frac{t-\lam_0\lam}{\tl} \right) \nonumber \\
&\times G\left( q_1\left( 3\alpha^2\frac{t-\lam_0\lam}{\tl} \right) \right) (\lam-\lam^2) \td{\lam} \nonumber \\
\approx & I_{e,0} \epss \left( 1+3\alpha^2\frac{t}{\tl} \right)^{2/3} G\left( q_1\left( 3\alpha^2\frac{t}{\tl} \right) \right) \nonumber \\
&\times \left[ 1-\frac{t}{\lam_0+t_c} \right] \label{eq:Le103} \eqd
\end{align}
\subsubsection{Calculations for $\epss<\epsu{c10}$ and $t_c>\lam_0$}
In this case not much needs to be done, since we can use the results calculated above. For the early times $t<\lam_0$ the light curve is well approximates by Eq. (\ref{eq:Le101}). Since for $t>\lam_0$ the retardation can be safely ignored, we find for $\lam_0<t<t_c$
\begin{align}
\Le =& 6 I_{e,0} \epss \intl_{0}^{1} \left( 1+3\alpha^2\frac{t-\lam_0\lam}{\tl} \right) \nonumber \\
&\times G\left( q_1\left( 3\alpha^2\frac{t-\lam_0\lam}{\tl} \right) \right) (\lam-\lam^2) \td{\lam} \nonumber \\
\approx & I_{e,0} \epss \left( 1+3\alpha^2\frac{t}{\tl} \right)^{2/3} G\left( q_1\left( 3\alpha^2\frac{t}{\tl} \right) \right) \label{eq:Le104} \eqd
\end{align}
For $t_c<t<t_c+\lam_0$ the result (\ref{eq:Le103}) is obviously valid.
\subsubsection{Calculations for $\epss>\epsu{c10}$ and $\epss<\epsu{10}$}
This case can only be achieved for $\epsu{10}>\epsu{c10}$, which translates to the requirement $t_c>\lam_0$. Then, $\tsu{10}>\lam_0$ and for $t<\lam_0$ one can approximate the light curve with Eq. (\ref{eq:Le101}). 

For times $\lam_0<t<\tsu{10}$ Eq. (\ref{eq:Le104}) can be used, while for $\tsu{10}<t<\tsu{10}+\lam_0$ one obtains
\begin{align}
\Le =& 6 I_{e,0} \epss \intl_{\frac{t-\tsu{10}}{\lam_0}}^{1} \left( 1+3\alpha^2\frac{t-\lam_0\lam}{\tl} \right) \nonumber \\
&\times G\left( q_1\left( 3\alpha^2\frac{t-\lam_0\lam}{\tl} \right) \right) (\lam-\lam^2) \td{\lam} \nonumber \\
\approx & I_{e,0} \epss \left( 1+3\alpha^2\frac{t}{\tl} \right)^{2/3} G\left( q_1\left( 3\alpha^2\frac{t}{\tl} \right) \right) \nonumber \\
&\times \left[ 1-\frac{t}{\lam_0+\tsu{10}} \right] \label{eq:Le105} \eqd
\end{align}
\subsubsection{Calculations for $\epss>\epsu{c10}$ and $\epss>\epsu{10}$}
Here $\tsu{10}<\lam_0$. For early times $t<\tsu{10}$ once more Eq. (\ref{eq:Le101}) represents the light curve. 

In the intermediate time domain $\tsu{10}<t<\lam_0$ the light curve is well approximated by Eq. (\ref{eq:Le102}), with the caveat that the upper limit in the integral $F_1(\epss)$ equals unity.

Lastly, for $\lam_0<t<\lam_0+\tsu{10}$ the light curve is well approximated by Eq. (\ref{eq:Le105}).
\subsection{Late time limit} \label{sec:ec10lt}
\subsubsection{Definitions} \label{sec:ec10ltdef}
For the late time limit $t>t_c$ some additional definitions are necessary. The second Heaviside function in Eq. (\ref{eq:Iec2}) gives the cut-off time
\begin{align}
\tsu{20} = \tl \left[ \frac{\gamma_0}{\gamma_{ec}} - \alpha_g \right] \label{eq:te20} \eqd
\end{align}
The condition $\tsu{20}<t_c$ actually yields the same energy condition as for the early time limit (Eq. (\ref{eq:eec10})). However, for $\tsu{20}<t_c$ the unretarded light curve cuts off before the late time limit even begins. Hence, only for energies $\epss<\epsu{c10}$ the late time limit contributes to the final light curve.

As in the respective SSC case in section \ref{sec:ssc10lt} for $\tsu{10}<\lam_0+t_c$ one finds
\begin{align}
\epss>\epsu{20} = \frac{\Gamec\gamma_0}{\left( \alpha+\frac{\lam_0}{\tl} \right) \left( \alpha+\frac{\lam_0}{\tl} + \Gamec \right)} \label{eq:ee20} \eqd
\end{align}
\subsubsection{Calculations for $\epss<\epsu{20}$}
Inserting Eq. (\ref{eq:Iec2}) into Eq. (\ref{eq:lc0}) we obtain for $t_c<t<\lam_0+t_c$
\begin{align}
\Le =& 6 I_{e,0} \epss \intl_{0}^{\frac{t-t_c}{\lam_0}} \left( \alpha_g+\frac{t-\lam_0\lam}{\tl} \right)^{2} G\left( q_2\left( \frac{t-\lam_0\lam}{\tl} \right) \right) \nonumber \\
& \times (\lam-\lam^2) \td{\lam} \nonumber \\
\approx & 3 I_{e,0} \alpha_g^2 \epss G\left( q_2(0) \right) \left( \frac{t-t_c}{\lam_0} \right)^2 \left[ 1-\frac{2}{3}\frac{t}{t_c+\lam_0} \right] \label{eq:Le201} \eqd
\end{align}
Here we used $(t-\lam_0\lam)/\tl\ll \alpha_g$ for a large part of the parameter space (which results in the approximation $\tau\rightarrow 0$ in $q_2(\tau)$ from Eq. (\ref{eq:q2})).

For times $\lam_0+t_c<t<\tsu{20}$ the retardation becomes unimportant, resulting in
\begin{align}
\Le =& 6 I_{e,0} \epss \intl_{0}^{1} \left( \alpha_g+\frac{t-\lam_0\lam}{\tl} \right)^{2} G\left( q_2\left( \frac{t-\lam_0\lam}{\tl} \right) \right) \nonumber \\
& \times (\lam-\lam^2) \td{\lam} \nonumber \\
\approx & I_{e,0} \epss \left( \alpha_g+\frac{t}{\tl} \right)^2 G\left( q_2\left( \frac{t}{\tl} \right) \right) \label{eq:Le202} \eqd
\end{align}

Lastly, the light curve for $\tsu{20}<t<\tsu{20}+\lam_0$ becomes
\begin{align}
\Le =& 6 I_{e,0} \epss \intl_{\frac{t-\tsu{20}}{\lam_0}}^{1} \left( \alpha_g+\frac{t-\lam_0\lam}{\tl} \right)^{2} G\left( q_2\left( \frac{t-\lam_0\lam}{\tl} \right) \right) \nonumber \\
& \times (\lam-\lam^2) \td{\lam} \nonumber \\
\approx & I_{e,0} \epss \left( \alpha_g + \frac{t}{\tl} \right)^2 G\left( q_2\left( \frac{t}{\tl} \right) \right) \nonumber \\
& \times \left[ 1-\frac{t}{\tsu{20}+\lam_0} \right] \label{eq:Le203} \eqd
\end{align}
\subsubsection{Calculations for $\epss>\epsu{20}$}
In this case $\tsu{20}<\lam_0+t_c$. Still, for $t_c<t<\tsu{20}$ we can use approximation (\ref{eq:Le201}).

The intermediate time domain $\tsu{20}<t<\lam_0+t_c$ is derived in appendix \ref{app:eca10b}, yielding
\begin{align}
\Le =& \frac{3}{8}I_{e,0} \Gamec^3 \epss F_2(\epss) \frac{\tl}{\lam_0} \left( \frac{t}{\lam_0} \right) \left[ 1-\frac{t}{\lam_0+t_c} \right] \label{eq:Le204} \eqd
\end{align}
As before, the explicit form of the integral function $F_2(\epss)$ is not necessary for the ongoing discussion, and we also find a break in the light curve at $t=\tsu{20}$.

Finally, for $t_c+\lam_0<t<\tsu{20}+\lam_0$ the result (\ref{eq:Le203}) can be used, too.
\subsection{Results for $\alpha\gg 1$}
Summarizing the calculations and stitching the results of the different time domains gives us the EC light curves for $\alpha\gg 1$. The results of the late time limit are only needed for low energies $\epss<\epsu{20}<\epsu{c10}$.

Beginning with the highest energies $\epss>\epsu{11}>\epsu{10}>\epsu{c10}$, the complete light curve becomes
\begin{align}
\Le = 3 I_{e,0} \epss G\left( q_1(0) \right) \frac{\left( \frac{t}{\lam_0} \right)^2}{1+\frac{t}{\tsu{10}}} \left[ 1-\frac{t}{\lam_0} \right] \label{eq:Le11} \eqd
\end{align}

For $\epss>\epsu{10}>\epsu{c10}$, but $\epss<\epsu{11}$ the light curve is significantly influenced by the rising part of the unretarded light curve. Thus,
\begin{align}
\Le =& 3 I_{e,0} \epss G\left( q_1(0) \right) \frac{\left( \frac{t}{\lam_0} \right)^2 \left( 1+3\alpha^2\frac{t}{\tl} \right)^{2/3}}{1+1.5\frac{t}{\tsu{10}}} \nonumber \\
& \times \left[ 1-\frac{t}{\lam_0} \right] \label{eq:Le12} \eqd
\end{align}
The factor $1.5$ in the denominator is a compromise. In principle this factor should equal unity, which gives excellent agreement for high $\epss$, while the agreement for low $\epss$ is not as good beyond the break. Using a factor two significantly improves the fit for low energies, but simultaneously decreases the goodness of the fit for high energies. With the factor $1.5$ both high and low energies are fit equally well.

For energies $\epsu{c10}<\epss<\epsu{10}$, which is only possible if $t_c>\lam_0$, the light curve becomes
\begin{align}
\Le =& 3 I_{e,0} \epss \frac{\left( \frac{t}{\lam_0} \right)^2}{1+3\left( \frac{t}{\lam_0} \right)^2} \left( 1+3\alpha^2\frac{t}{\tl} \right)^{2/3} \nonumber \\
& \times G\left( q_1\left( 3\alpha^2\frac{t}{\tl} \right) \right) \label{eq:Le13} \eqd
\end{align}

In the energy domain $\epss<\epsu{c10}$ we find for $\epss>\epsu{20}$
\begin{align}
\Le =& 3 I_{e,0} \epss G\left( q_1(0) \right) \left( \frac{t}{\lam_0} \right)^2 \left( 1+3\alpha^2\frac{t}{\tl} \right)^{2/3} \nonumber \\
& \times \left[ 1-\frac{t}{t_c+\lam_0} \right] \label{eq:Le14} \eqd
\end{align}

In the last energy domain $\epss<\epsu{20}<\epsu{c10}$ both early and late time limit are needed. Similar to the SSC case, we cannot stitch them properly at $t=t_c$, since the approximation during the derivation of the electron energy distribution (c.f. Schlickeiser et al., 2010) causes a sharp upturn in the light curve, which cannot be covered by stitched breaks. Thus,
\begin{align}
\Le =& 3 I_{e,0} \epss \frac{\left( \frac{t}{\lam_0} \right)^2}{1+3\left( \frac{t}{\lam_0} \right)^2} \nonumber \\
& \times \left\{ \HF{t_c-t} \left( 1+3\alpha^2\frac{t}{\tl} \right)^{2/3} G\left( q_1\left( 3\alpha^2 \frac{t}{\tl} \right) \right) \right. \nonumber \\
& + \left. \HF{t-t_c} \left( \alpha_g+\frac{t}{\tl} \right)^2 G\left( q_2\left( \frac{t}{\tl} \right) \right)  \right\} \label{eq:Le15} \eqd
\end{align}

We already discussed some aspects above. Some lightcurves fit the numerical result better than others, but the overall agreement is encouraging. Plots and a detailed discussion are given in section \ref{sec:dis}.
%
%
\section{Discussion} \label{sec:dis}
\begin{figure*}
\begin{minipage}{0.49\linewidth}
\centering \resizebox{\hsize}{!}
{\includegraphics{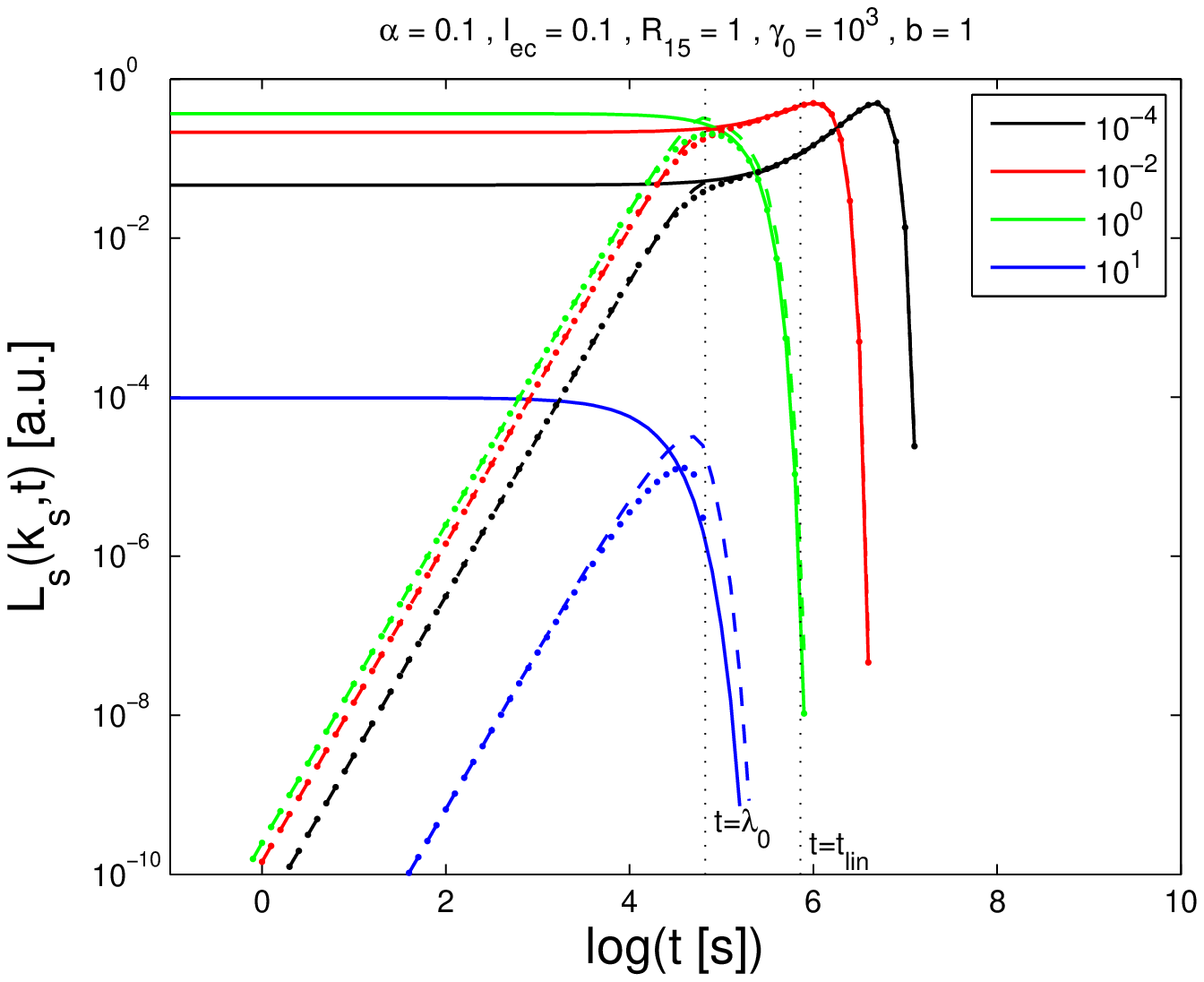}}
\end{minipage}
\hspace{\fill}
\begin{minipage}{0.49\linewidth}
\centering \resizebox{\hsize}{!}
{\includegraphics{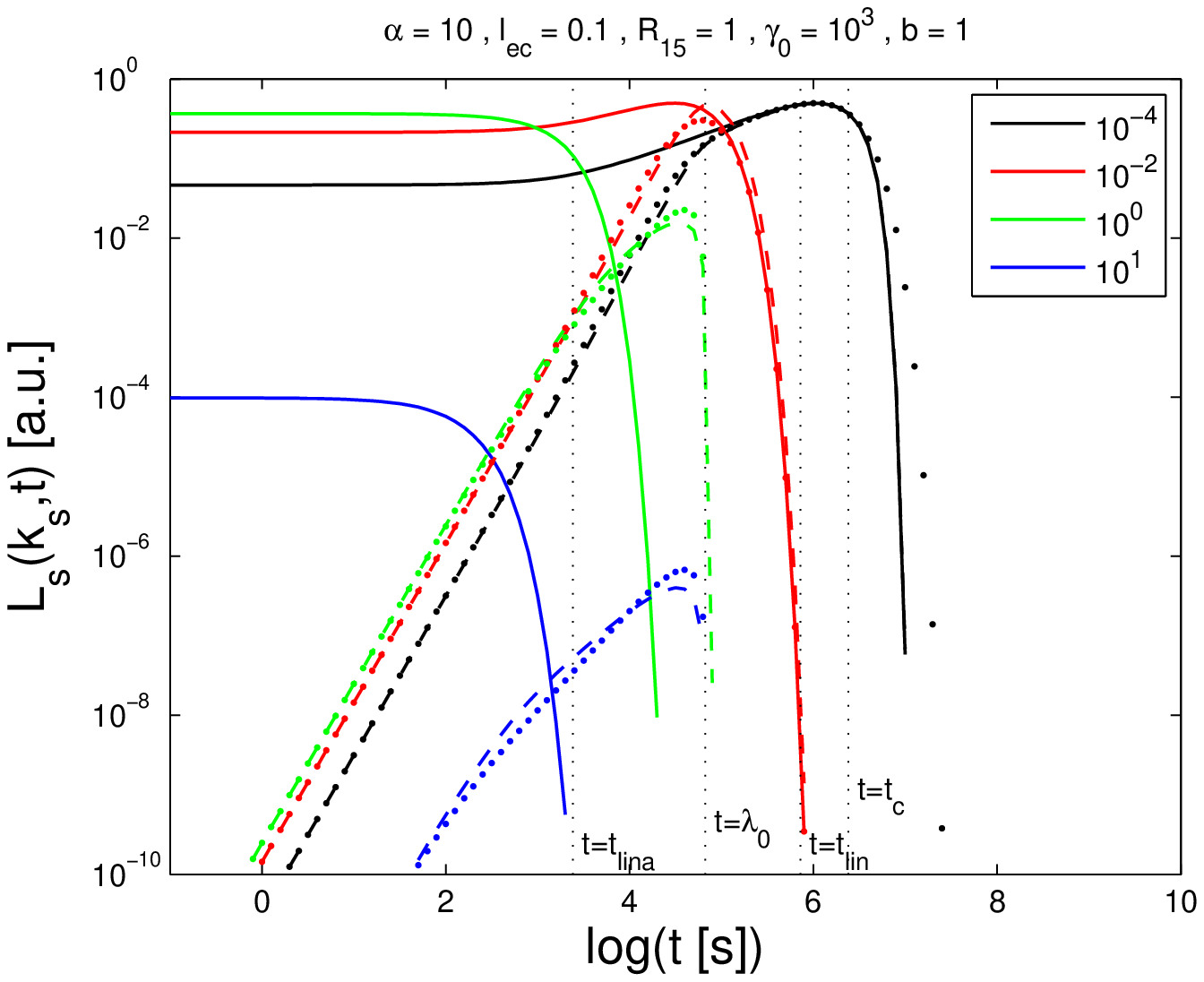}}
\end{minipage}
\newline
\begin{minipage}{0.49\linewidth}
\centering \resizebox{\hsize}{!}
{\includegraphics{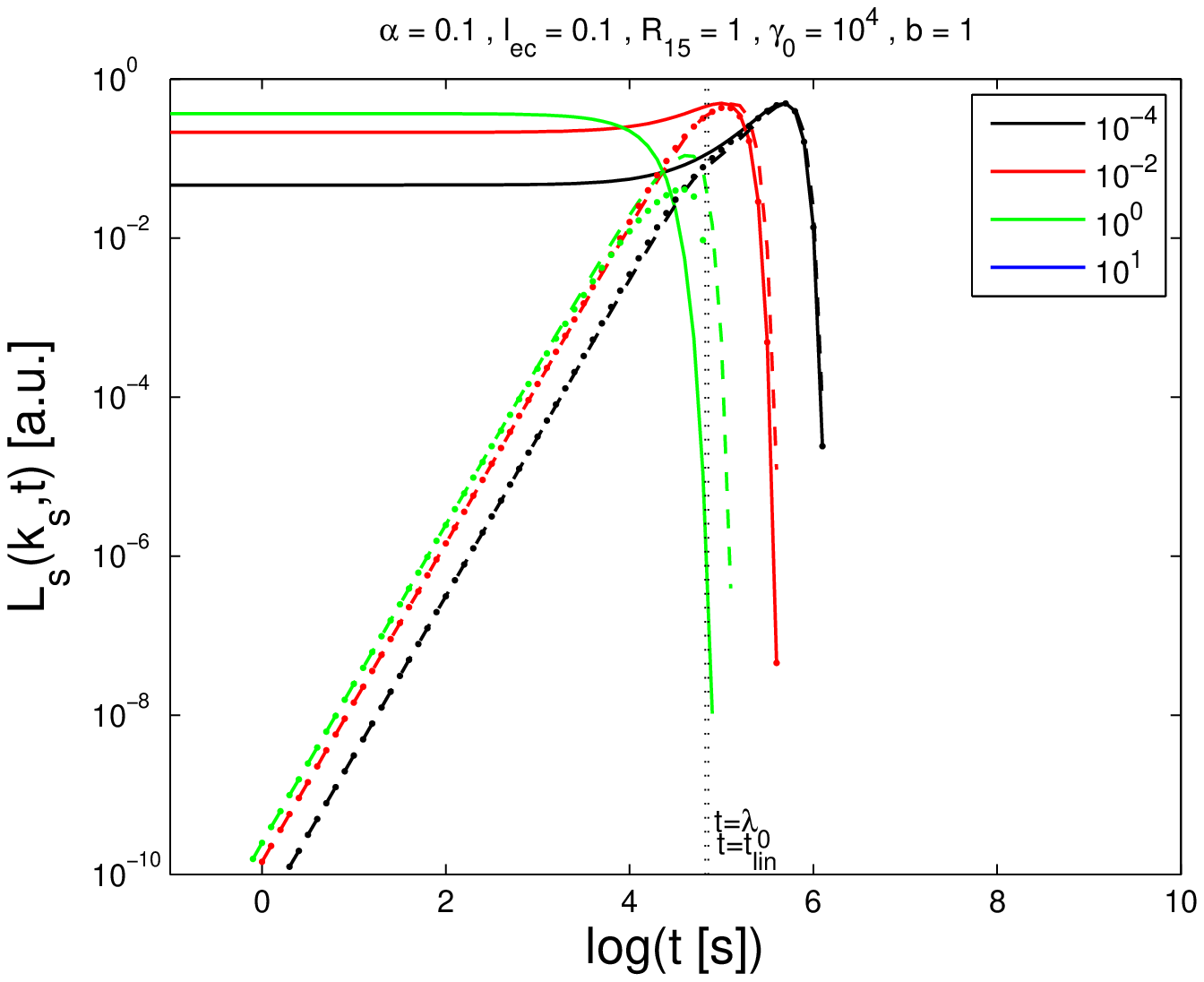}}
\end{minipage}
\hspace{\fill}
\begin{minipage}{0.49\linewidth}
\centering \resizebox{\hsize}{!}
{\includegraphics{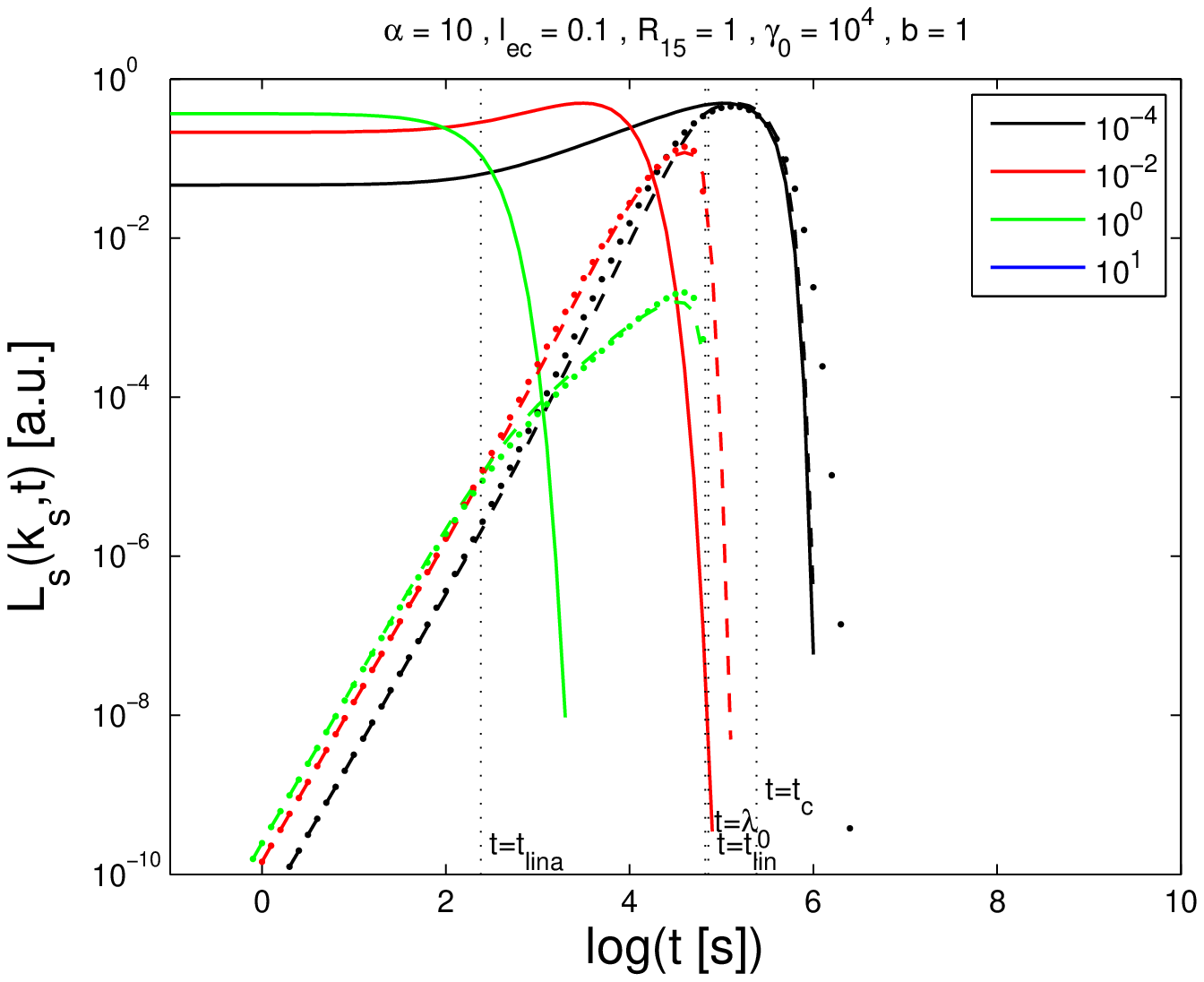}}
\end{minipage}
\caption{Unretarded SSC light curves (full), and analytical (dotted) and numerical (dashed) retarded SSC light curves over a logarithmic time axis. The parameters are given at the top and values of $k_s$ in the legend. Note: The light curves are devided by $I_{s,0}$, and $t_{lina} = \tl/3\alpha^2$.}
\label{fig:Lslog}
\end{figure*} 
\begin{figure*}
\begin{minipage}{0.49\linewidth}
\centering \resizebox{\hsize}{!}
{\includegraphics{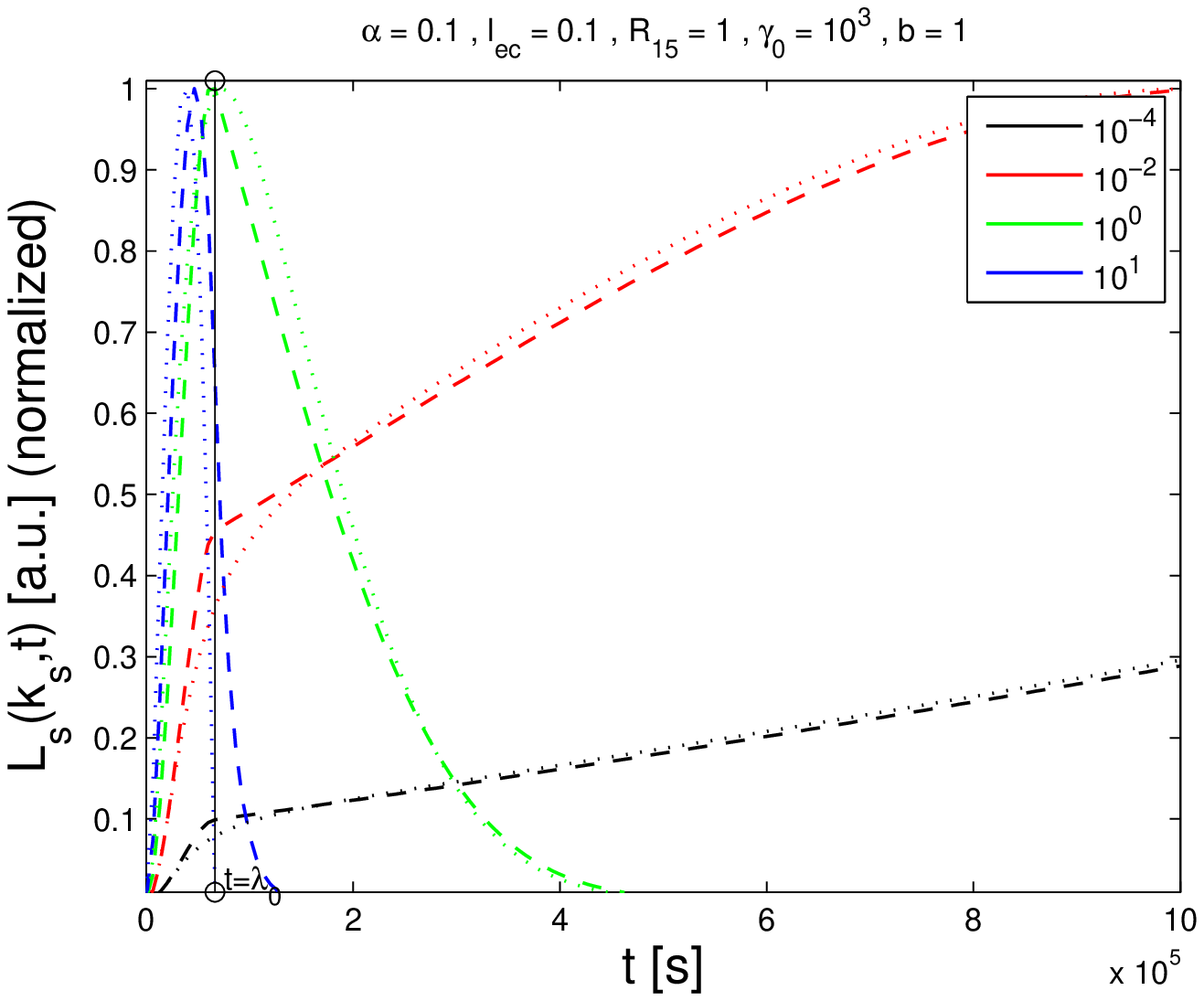}}
\end{minipage}
\hspace{\fill}
\begin{minipage}{0.49\linewidth}
\centering \resizebox{\hsize}{!}
{\includegraphics{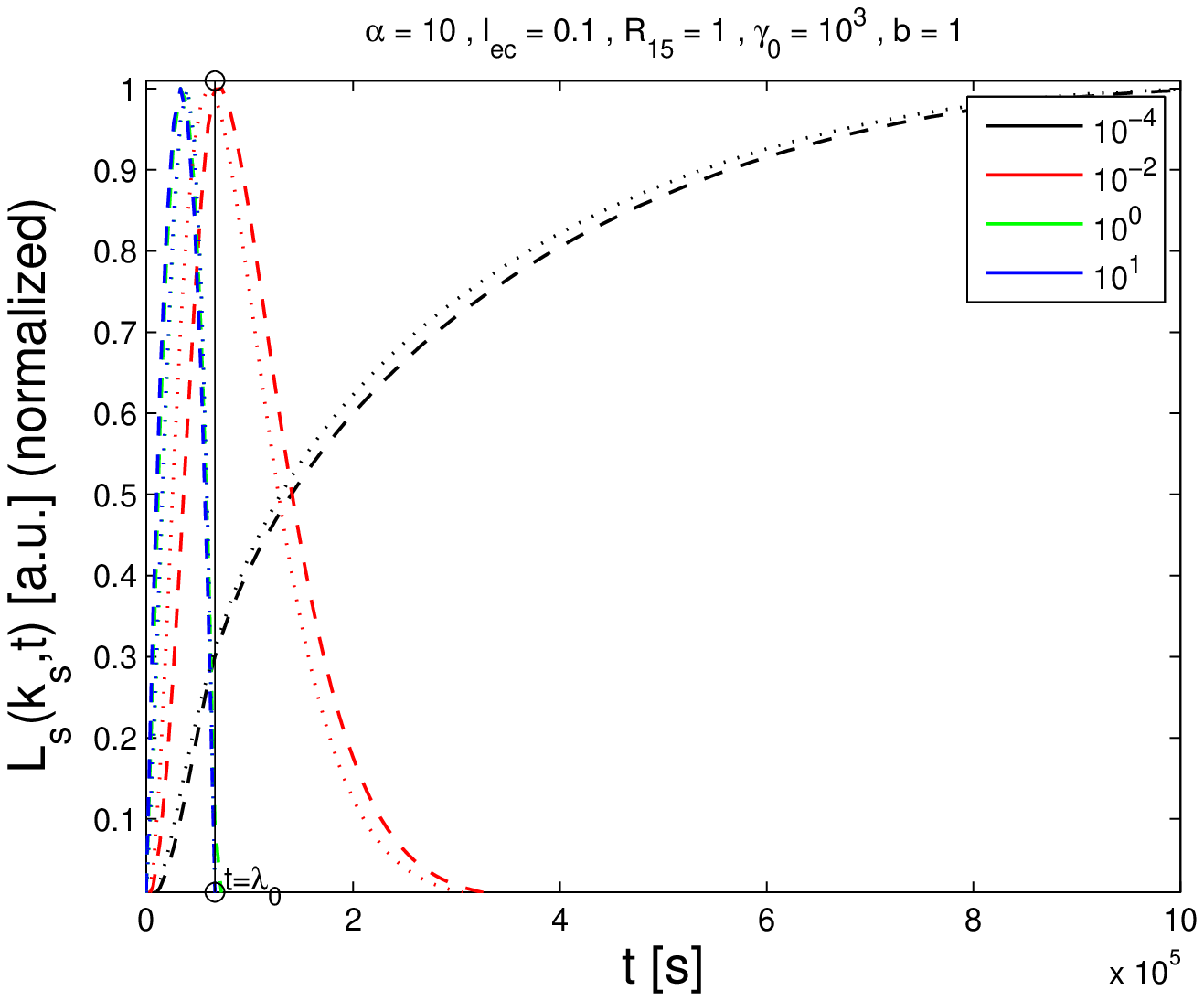}}
\end{minipage}
\newline
\begin{minipage}{0.49\linewidth}
\centering \resizebox{\hsize}{!}
{\includegraphics{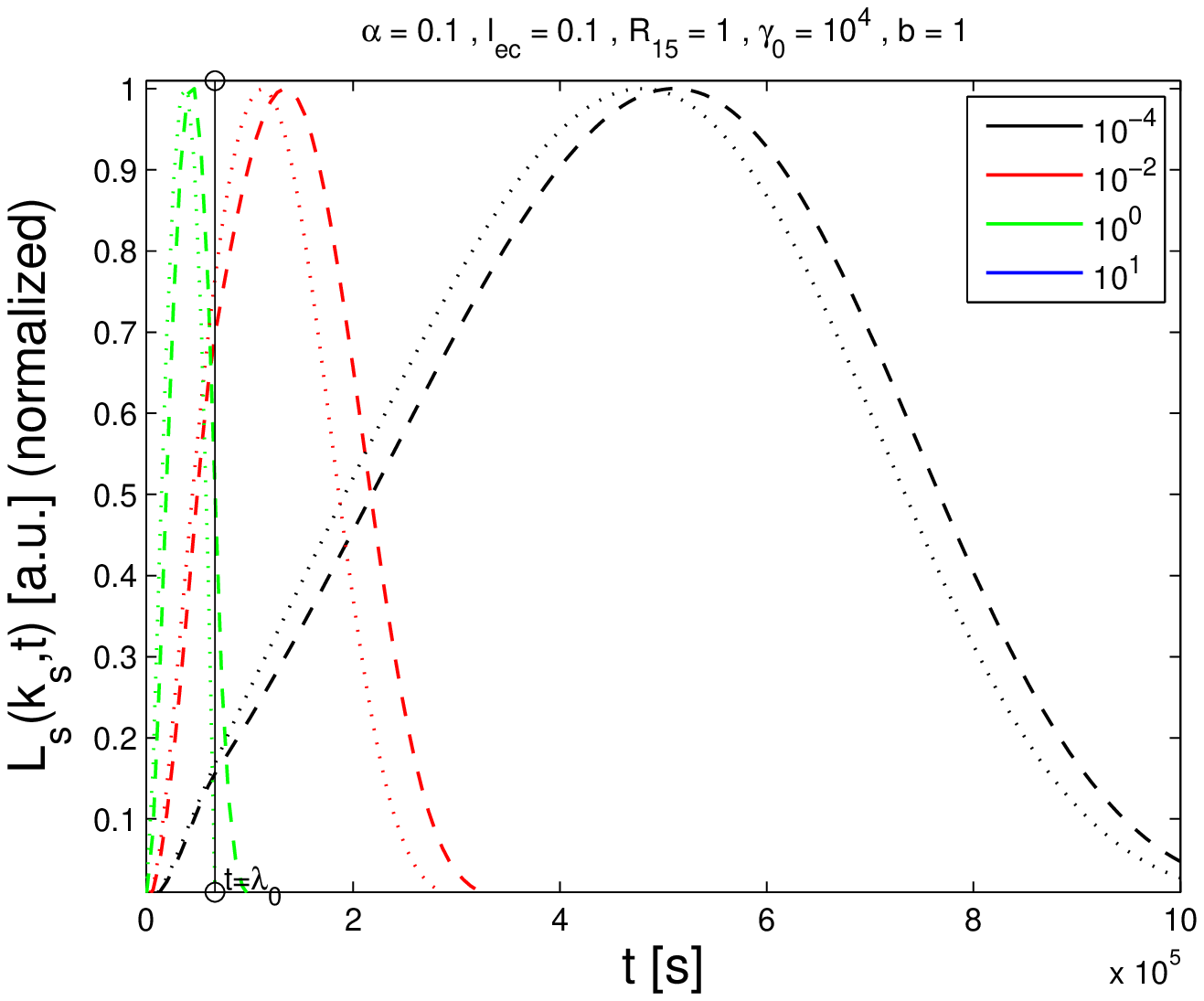}}
\end{minipage}
\hspace{\fill}
\begin{minipage}{0.49\linewidth}
\centering \resizebox{\hsize}{!}
{\includegraphics{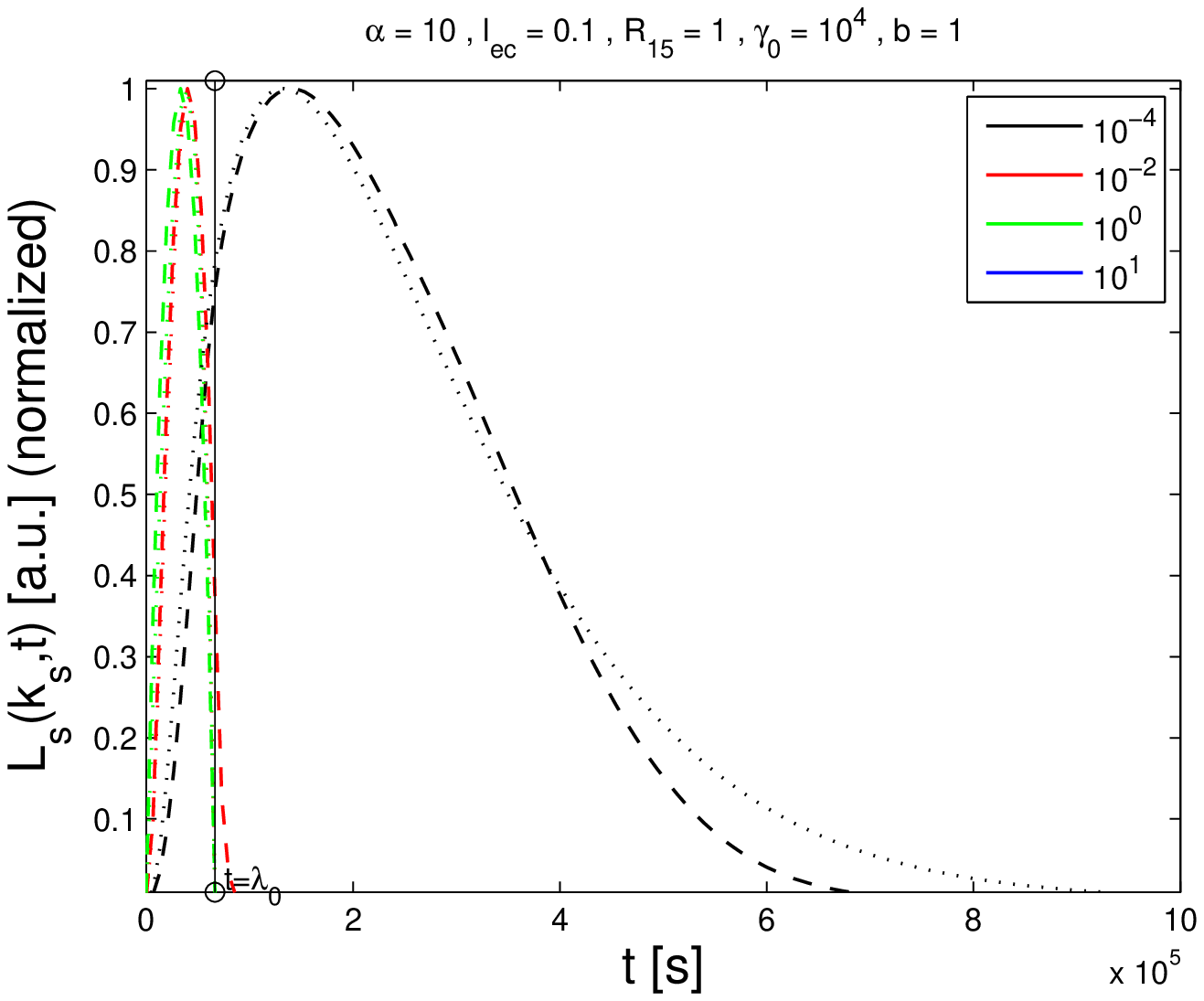}}
\end{minipage}
\caption{Normalized analytical (dotted) and numerical (dashed) retarded SSC light curves over a linear time axis. The parameters are given at the top and values of $k_s$ in the legend. The vertical line marks the LCT $\lam_0$ and the horizontal range is $15\lam_0$.}
\label{fig:Lslin}
\end{figure*} 
%
%
\begin{figure*}
\begin{minipage}{0.49\linewidth}
\centering \resizebox{\hsize}{!}
{\includegraphics{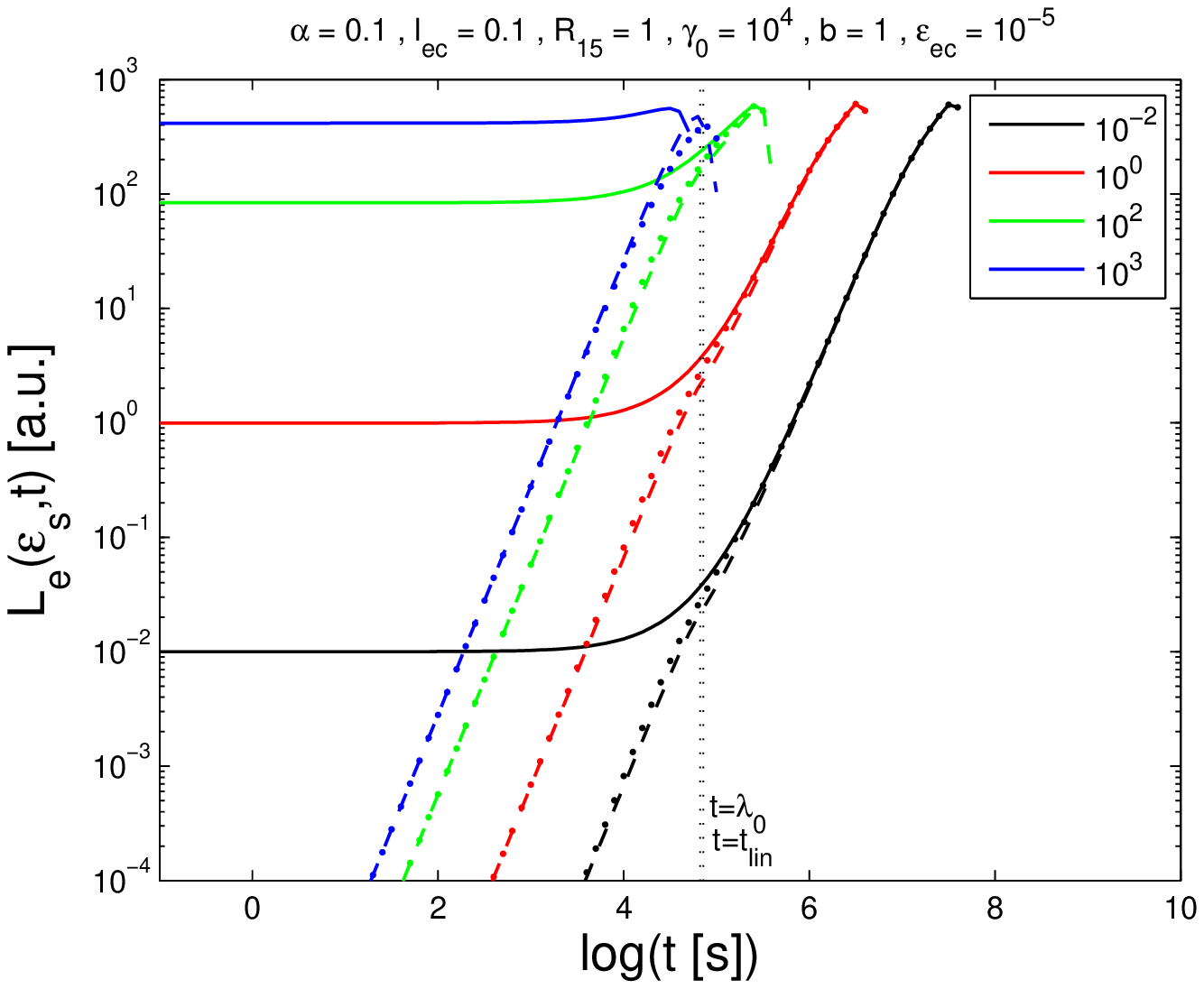}}
\end{minipage}
\hspace{\fill}
\begin{minipage}{0.49\linewidth}
\centering \resizebox{\hsize}{!}
{\includegraphics{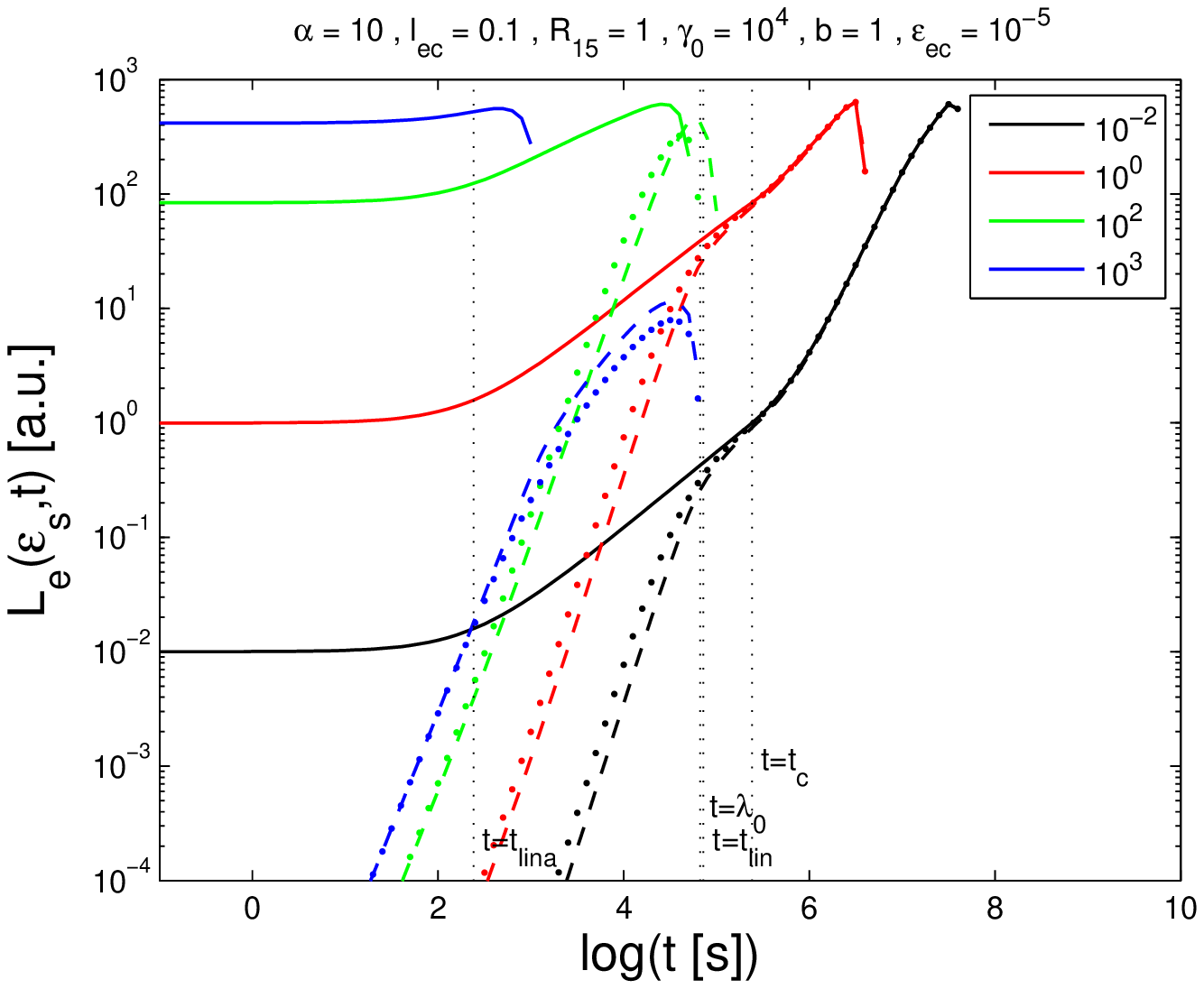}}
\end{minipage}
\newline
\begin{minipage}{0.49\linewidth}
\centering \resizebox{\hsize}{!}
{\includegraphics{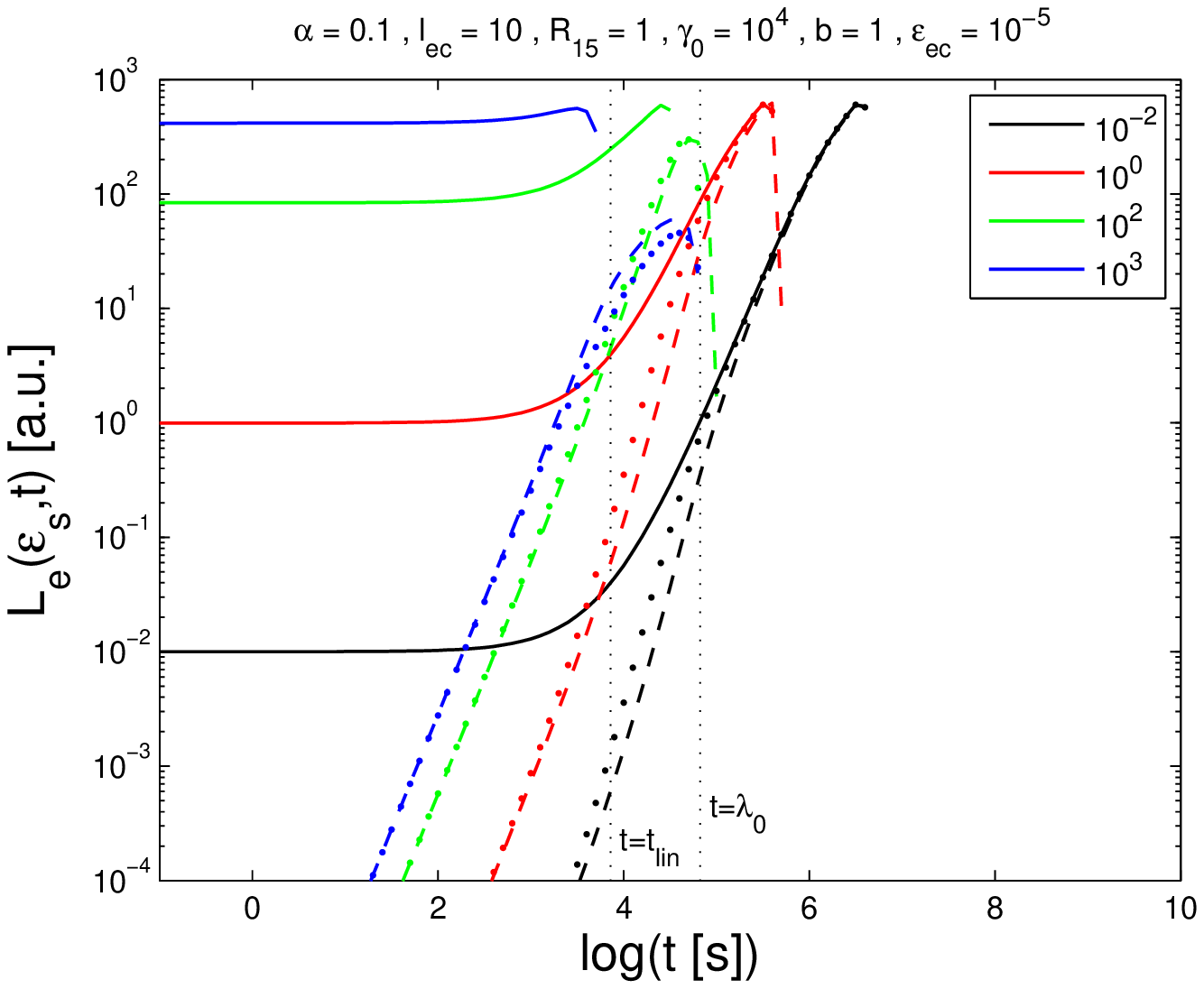}}
\end{minipage}
\hspace{\fill}
\begin{minipage}{0.49\linewidth}
\centering \resizebox{\hsize}{!}
{\includegraphics{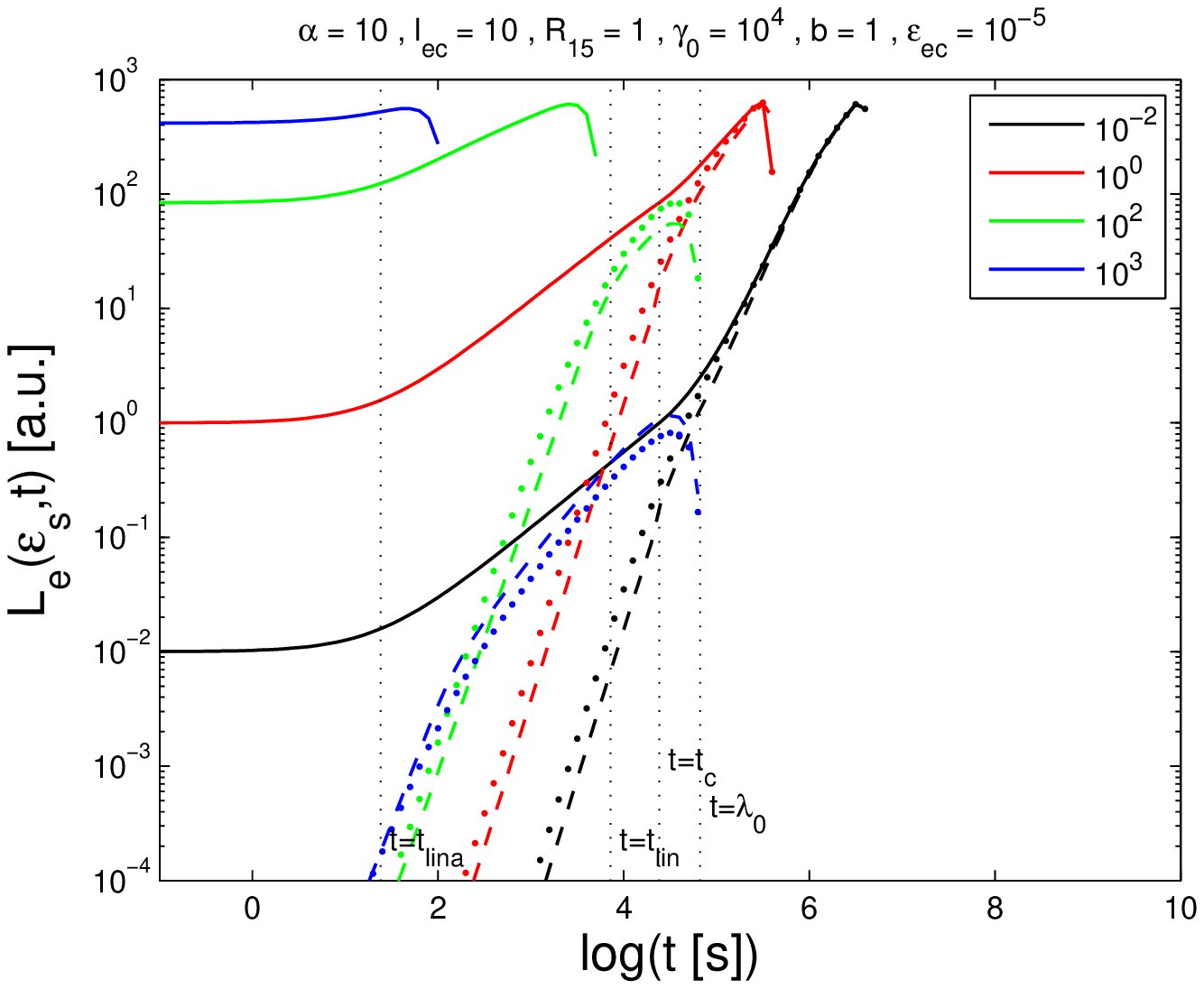}}
\end{minipage}
\caption{Unretarded EC light curves (full), and analytical (dotted) and numerical (dashed) retarded EC light curves over a logarithmic time axis. The parameters are given at the top and values of $\epsilon_s$ in the legend. Note: The light curves are devided by $I_{e,0}$, and $t_{lina} = \tl/3\alpha^2$.}
\label{fig:Lelog}
\end{figure*} 
\begin{figure*}
\begin{minipage}{0.49\linewidth}
\centering \resizebox{\hsize}{!}
{\includegraphics{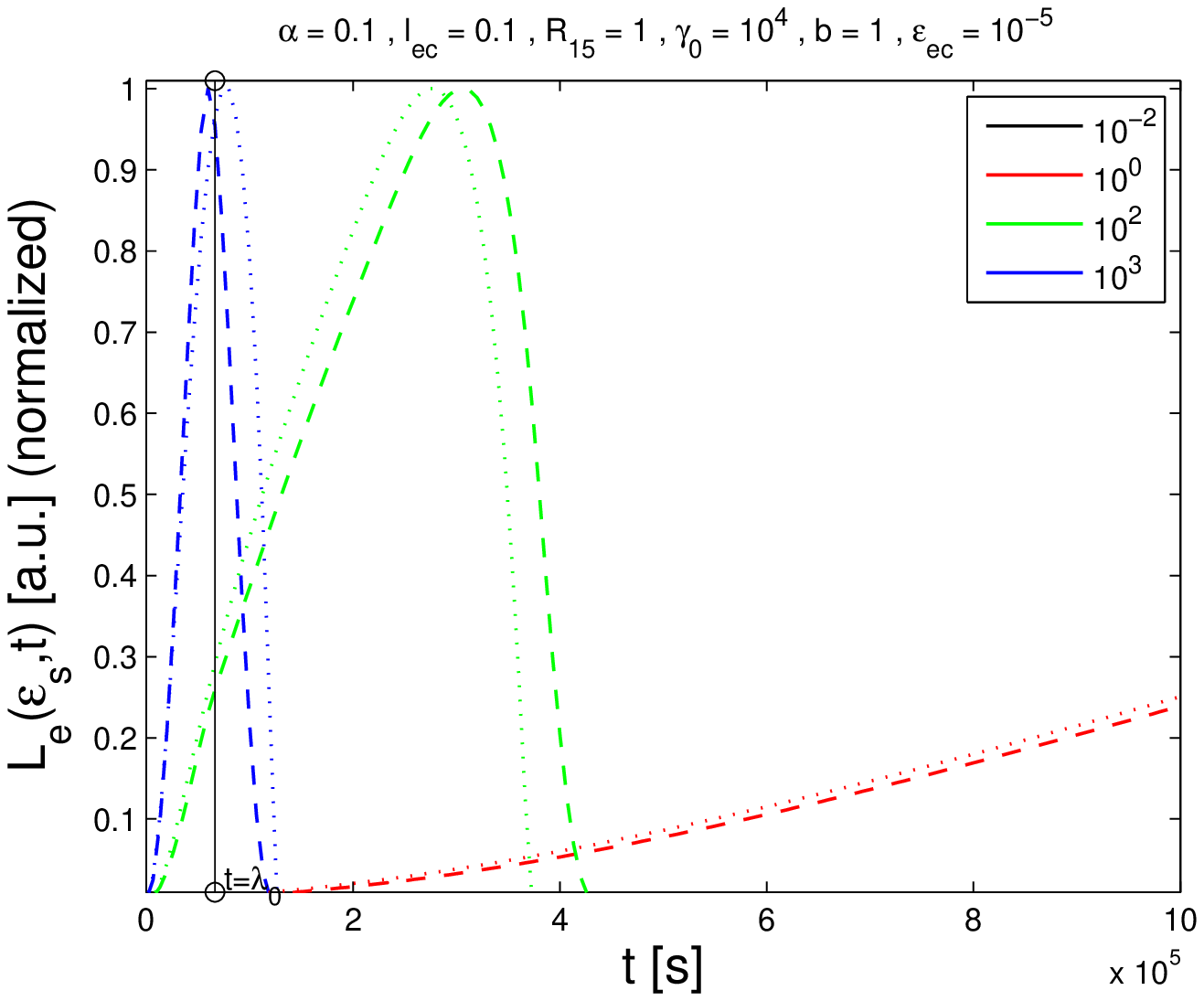}}
\end{minipage}
\hspace{\fill}
\begin{minipage}{0.49\linewidth}
\centering \resizebox{\hsize}{!}
{\includegraphics{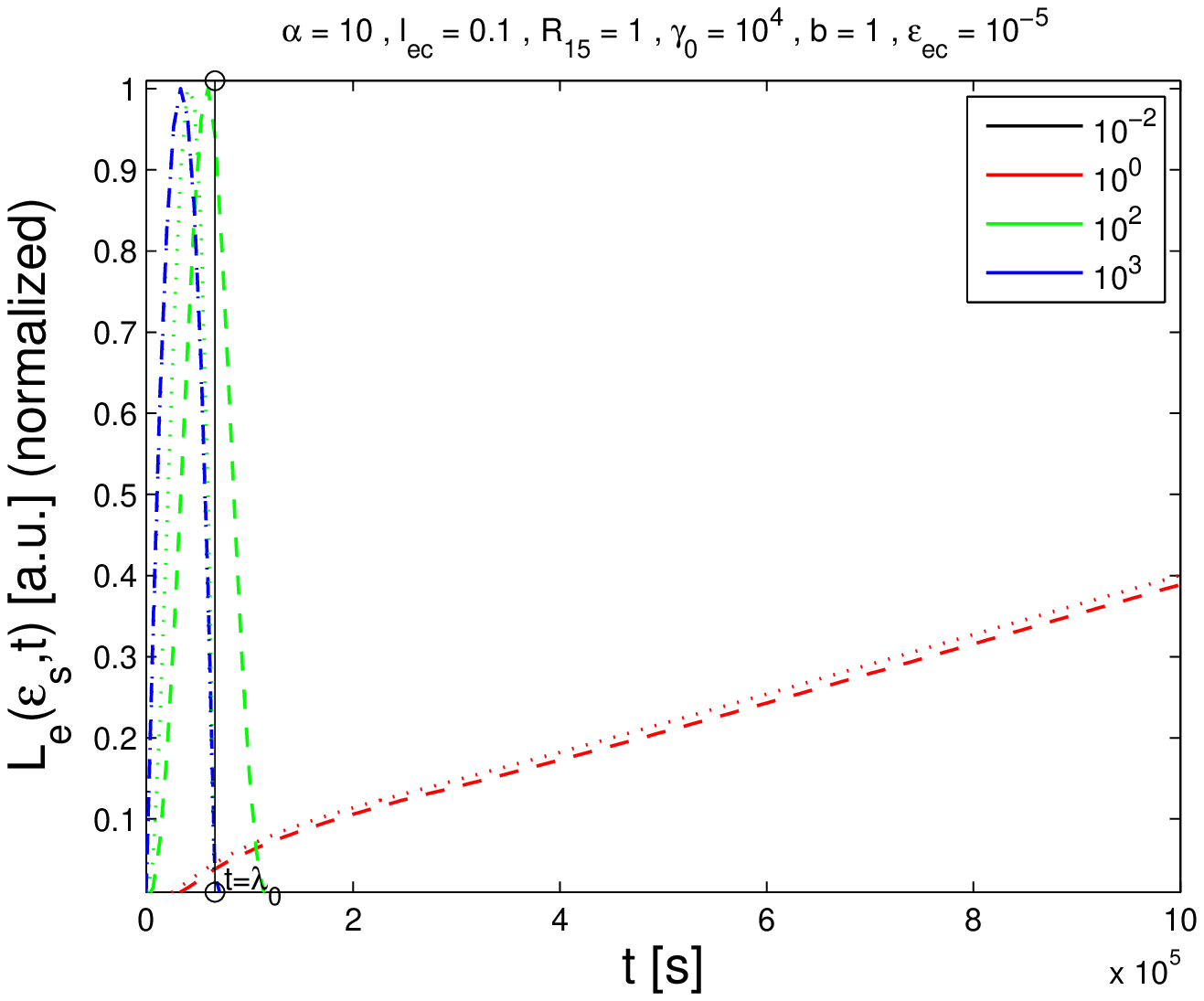}}
\end{minipage}
\newline
\begin{minipage}{0.49\linewidth}
\centering \resizebox{\hsize}{!}
{\includegraphics{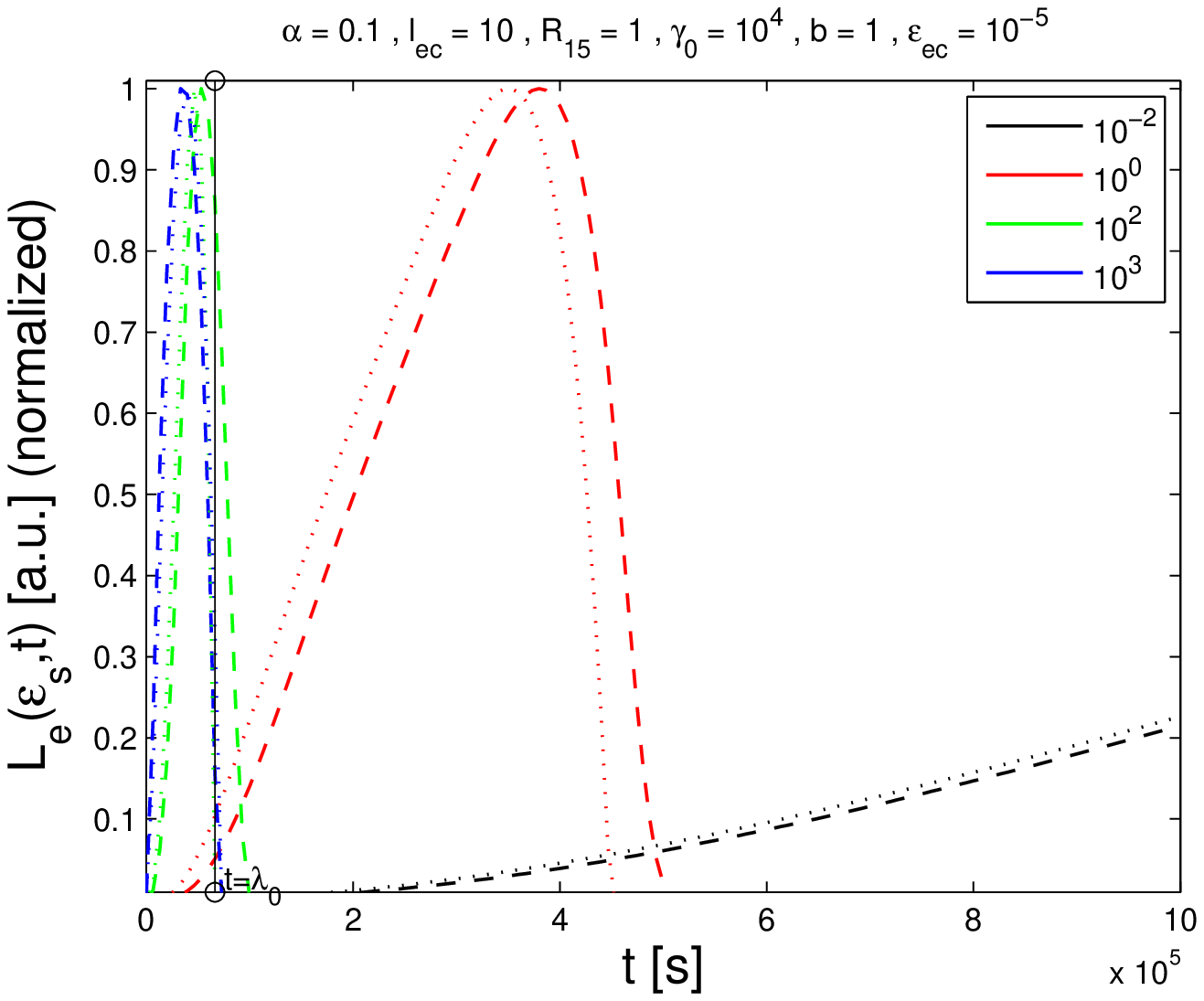}}
\end{minipage}
\hspace{\fill}
\begin{minipage}{0.49\linewidth}
\centering \resizebox{\hsize}{!}
{\includegraphics{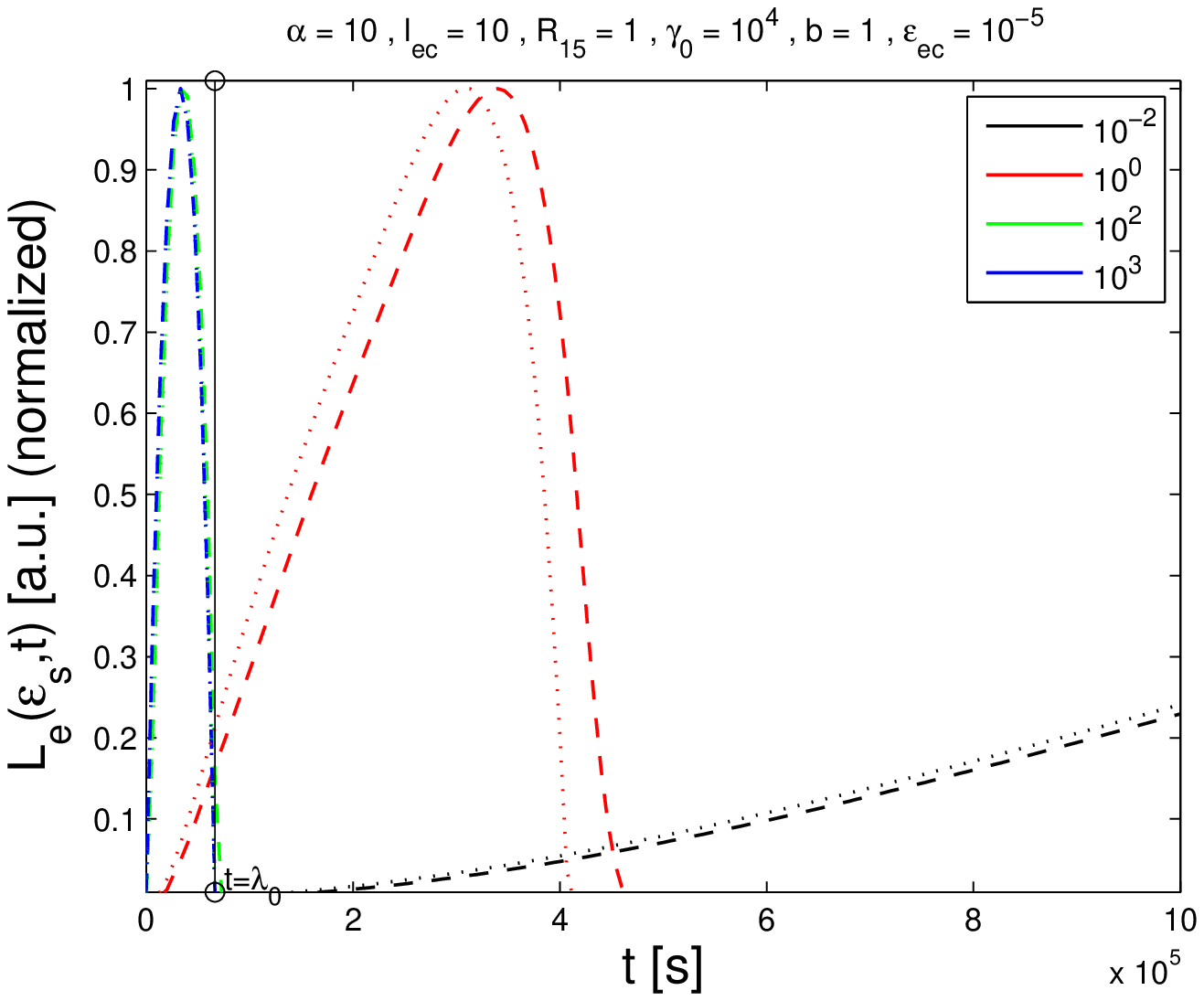}}
\end{minipage}
\caption{Normalized analytical (dotted) and numerical (dashed) retarded EC light curves over a linear time axis. The parameters are given at the top and values of $\epsilon_s$ in the legend. The vertical line marks the LCT $\lam_0$ and the horizontal range is $15\lam_0$.}
\label{fig:Lelin}
\end{figure*} 
%
%
In Figs. \ref{fig:Lslog} - \ref{fig:Lelin} we present model light curves for the {analytical results of SSC and EC emission (dotted lines)}. We show two versions each, where the first version plots the light curves over a logarithmic axis, while in the second version the light curves are shown over a linear time axis, respectively. The parameters are the same in both versions. Light curves are normally displayed over linear time-axis for obvious reasons. In our theoretical discussion this has the disadvantage that we can only present a very narrow part of the light curves in a linear plot. Logarithmic plots circumvent this disadvantage. On the other hand, the early phases of the light curves cover a large part of the plot and may seem overly enhanced compared to the later parts. The rising phase of the light curves might in this case {appear to be} more important than the behaviour around the maxima. The following description of plot properties focuses on the logarithmic plots. In most cases they also apply for the linear plots.

It might seem difficult to directly compare the theoretical light curves with observational ones, since in observations it is hard to mark the starting point $t=0$. However, the starting point is not really important, since one can always try to fit the maxima by shifting the theoretical light curves. As noted earlier, it is not our purpose to fit the theoretical light curves to data, since the injection model is too simple for that. This work is meant to highlight the differences between the linear and non-linear cooling cases, and that the latter might be important to model rapid flares of blazars.

{In order to check the validity of the analytical approximations, we performed numerical integrations of Eq. (\ref{eq:lc0}) for the respective cases using a Gauss-Kronrod quadrature method. The numerical curves are shown as dashed lines in Figs. \ref{fig:Lslog} - \ref{fig:Lelin}.

The full curves in Figs. \ref{fig:Lslog} and \ref{fig:Lelog} are the unretarded light curves, where the retardation is neglected. They are not shown in the linear plots, since they focus on the more important details of the analytical and numerical curves.}

In the following, we first describe the SSC and EC light curves, respectively, and then discuss some general considerations that apply for all cases, including synchrotron.
\subsection{The SSC light curves}
The logarithmic and linear SSC light curves are presented in Figs. \ref{fig:Lslog} and \ref{fig:Lslin}, respectively. The general points, which were briefly mentioned during the analytical calculations, are obvious.

For times smaller than the LCT the light curves increase, because the emission of an increasing number of slices becomes observable. At first, the increase of the emission is quadratic for all energies. If the unretarded light curve cuts off before the LCT, the light curves exhibit a break by unity at the time of the {unretarded} cut-off, and the retarded light curves are {cut off at the LCT.} This is reasonable, since at later times (between the unretarded cut-off and the LCT) only a decreasing number of slices in the back contribute. Light curves of lower energies, where the unretarded cut-off is beyond the LCT, break at the LCT exhibiting the behaviour of the unretarded light curve {for later times}. 

As a matter of fact, the behaviour of the light curves before the LCT depends both on the retardation and the geometry of the source. The former causes the increase, while the latter controls the power of the increase. In the spherical geometry used in this work the main increase is quadratic, while for example in a cylindrical geometry the main increase would be linear. Furthermore, the early part before the LCT is also influenced by the cooling time $\tl$, if the unretarded light curve cuts off beyond the cooling time, and if the cooling time is shorter than the LCT. For example the red and the black curve in the top right plot of Fig. \ref{fig:Lslog} deviate from a quadratic increase between the cooling time for large $\alpha$ ($t_{lina} = \tl/3\alpha^2$) and the LCT. A similar behaviour is obvious for the black curves in both lower plots in Fig. \ref{fig:Lslog}.

Such features might already be used to discriminate between different models and the parameters of the source. On the other hand, such small deviations from a simple power-law might be hard to detect, since very precise sampling is necessary to confirm the powers and the breaks. Current X-ray and $\gamma$-ray observatories are not able to measure with the necessary precision.

The behaviour of the retarded light curves beyond the LCT matches, as stated before, the unretarded one quite well, which is expected, since the retardation becomes increasingly unimportant. However, as is obvious in Fig. \ref{fig:Lslin}, the light curves are still influenced a bit by the retardation, since the numerical curve is slightly delayed compared to the analytical one. This is reasonable, since a small delay will always be present due to the finite extension of the emission region. This is most obvious in the linear plots, which cover only a narrow time range around the LCT. The analytical results leave this retardation effect aside. The delay of the exponential cut-off of the black analytical curve in the right plots of Fig. \ref{fig:Lslog} is due to the approximation to use the exponential cut-off of the early time limit ($t<t_c$) also beyond the crossover time $t_c$. Since the exponential cut-off of the early time limit is less severe than the one of the late time limit, the analytical curve cuts off slower than the numerical and the unretarded curve.

Comparing the plots for low (left plots) and high (right plots) values of the injection parameter $\alpha$ clear differences become obvious. First of all, since the cooling time $t_{lina}$ is much reduced in the $\alpha>1$ case, the unretarded light curves are variable on shorter time scales than the light curves for $\alpha<1$. The variability time scales of the unretarded light curve have direct consequences on the retarded light curves {as discussed above}. Thus, only by increasing $\alpha$, the behaviour of the light curves is completely altered. For example in the top plots of Fig. \ref{fig:Lslog}, the blue curve exhibits the break at earlier times (that is to say, after a shorter flare duration), the green curve cuts off at the LCT, the exponential cut-off of the red curve takes place an order of magnitude earlier, and {temporal shape of the black curve exhibits a clear deviation form the $\alpha\ll 1$ case}. These examples show that the non-linear, time-dependent SSC cooling significantly shortens and alters the flare duration while most of the input parameters are unchanged. Such short temporal variability would normally be explained with higher electron energies, which also shortens the variability time (see lower plots on Fig. \ref{fig:Lslog}), but which would not change the temporal appearance apart from retardation effects (c.f. left plots in Fig. \ref{fig:Lslog}).
\subsection{The EC light curves}
Figs. \ref{fig:Lelog} and \ref{fig:Lelin} display the logarithmic and linear plots of the EC light curves, respectively. In order to decrease the variability time from the top plots to the lower ones we increase the external Compton parameter $l_{ec}$ instead of the electron Lorentz factor $\gamma_0$.

Interestingly, apart from specific temporal shapes of the light curves, the behaviour of the EC light curves are remarkably similar to the SSC case. Before the LCT the light curves are dominated by the retardation and the geometry, while the cut-off of the unretarded light curves might cause a break in the retarded light curve. Beyond the LCT the shape of the unretarded light curves is recovered, although the mentioned delay of the numerical light curve is also evident here (especially in Fig. \ref{fig:Lelin}).

The sharp cut-off of the light curves is due to the soft photon field used for the inverse Compton scattering. As mentioned in section \ref{sec:app}, the soft external photon field is assumed to be monochromatic with normalized energy $\epec${.\footnote{In the plots we used $\epec=10^{-5}$, which is in the UV range near the Ly$_{\alpha}$-line, implying the inverse Compton process to be in the Thompson limit}} Thus, one can expect a different temporal shape of the EC light curve compared to the SSC case, where the broad (non-thermal) synchrotron emission is scattered.
\subsection{General remarks}
Having discussed the temporal shapes of the SSC and EC light curves separately, we now focus on some general aspects, which are even true for the synchrotron light curves calculated by ZS.

It is a general feature of all cases that for high energies, where the unretarded light curve cuts off long before the LCT, the retarded light curves exhibit lower maximum fluxes than the unretarded light curves. However, we expect photon number conservation, implying that the same number of photons is emitted by the source (this might be different when taking into account photon-photon absorption). The retardation merely causes the photons to be observed over longer time scales, since the slices contribute at different times to the observed emission, while the overall produced number of photons is the same as without retardation. Since for low energies the maximum of the light curve, and thus the bulk of photon emission is beyond the LCT the photon numbers should also be conserved.

The delay due to retardation has consequences also for the observed spectral energy distributions (SEDs), since snapshots of the SED at different times will show a flux at an energy where, if retardation is not taken into account, no flux should be detected. The total SEDs (that is, integrated over the entire flare) should therefore not be affected by the retardation.

The earlier cut-off of the high energies for high $\alpha$ in the unretarded light curves causes the retarded light curves to exhibit lower photon fluxes compared to the low $\alpha$ case. This flux reduction for high energies can also be seen in the SEDs, where an additional break appears, which strongly depends on the value of $\alpha$ (see Schlickeiser et al., 2010; Zacharias \& Schlickeiser, 2012a; 2012b). Both breaks in the SED and flux reduction in the light curve are due to more intense electron energy losses of the non-linear, time-dependent SSC cooling.

The light curves presented in this theoretical investigation have some features that are interesting for observational purposes. If multi-wavelength light curves are available, the combination should give some clues regarding the parameters of the source. Since both SED and light curves must in principle be fit with the same set of parameters, the light curves serve as a test for the parameters usually deduced from SED modelling. The variability time scale observed in the light curves is generally equalled with the LCT. This is true for higher energies, and can be discriminated if some of the higher energy light curves peak at roughly the same time. The maximum of lower energy light curves will occur at increasingly later times. However, this also depends on the injection energy of the electrons. Hence, if most light curves are close to the LCT, the initial electron energy is high, while for light curves peaking significantly later than the LCT the electron energy is low.

This interpretation is, however, not directly applicable, since the non-linear SSC cooling might decrease the variability times, too. As stated above, well sampled light curves might be used to discriminate between the high and low $\alpha$ cases, since the temporal shape of the light curves is altered. That this is difficult with current instruments and observations {has been stated} above as well, but might be less an issue if very detailed light curves with very precise time binning should become available.

The LCT can be further deduced from the symmetry of the light curves around their maximum. Without going into details, the light curves cutting off at the LCT are symmetric around the maximum, meaning that the times from half the maximum to the maximum and down again to half the maximum are roughly equally long. For the synchrotron and the SSC light curves for maxima occurring beyond the LCT the decay time is longer than the rising time. Hence, in such cases the light curves are skewed towards earlier times (c.f. Fig. \ref{fig:Lslin}). For the EC light curve, most likely due to the simple assumption of line-like soft target photons, the decay time is shorter than the rising time, and the light curves are skewed towards later times (c.f. Fig. \ref{fig:Lelin}). Thus, a symmetric light curve implies that the maximum is close to the LCT, while an asymmetric light curve belongs to a maximum beyond the LCT.

To conclude, the theoretical investigations outlined in this paper have clearly shown, again, that rapid blazar flares might be explained by the non-linear, time-dependent SSC process, and that significantly different features arise in the SSC and EC light curves compared to the purely linear cooling case. Explaining very short flares with a non-equilibrium, single injection model seems a viable alternative to the standard equilibrium models. Furthermore, with the assumption of very small emission regions, the non-linear SSC process would be favoured over the linear-only models (c.f. Eq. (\ref{eq:alpha}) {of this work}, and Eq. (19) of Schlickeiser et al., 2010). 

A forthcoming paper shall deal with the aspects of photon-photon absorption, which has been neglected so far in the discussions.
%
%
\section*{Acknowledgements}
The author thanks the referee for a careful reading of the manuscript and valuable suggestions for its improvement. \\
Discussions with R. Schlickeiser and G. Cologna, as well as the support from the German Ministry for Education and Research (BMBF) through the Verbundforschung Astroteilchenphysik grant 05A11VH2 are gratefully acknowledged.
%
%
\appendix
\section{Calculation of the intensities} \label{app:int}
Here we summarize the basic calculations leading to the SSC and EC intensities. Details can be found in Zacharias \& Schlickeiser (2012a; 2012b).

The basic formula to calculate isotropic intensities of process $i$ is
\begin{align}
I_i(\epss,t) = \frac{R}{4\pi} \intl_0^{\infty} n(\gamma,t) p_i(\epss,\gamma) \td{\gamma} \eqd \label{eq:appI}
\end{align}

The volume-averaged electron number density $n(\gamma,t)$ can be calculated from the kinetic equation (Kardashev, 1962)
\begin{align}
\frac{\pd{n(\gamma,t)}}{\pd{t}} - \frac{\pd{}}{\pd{\gamma}} \left[ |\dot{\gamma}(\gamma,t)|_{tot} n(\gamma,t) \right] = S(\gamma,t) \label{eq:appke} \eqd
\end{align}
The injection term is chosen to be an instantaneous burst of monochromatic highly relativistic electrons: $S(\gamma,t) = q_0 \DF{\gamma-\gamma_0} \DF{t}$. The electron cooling consists of linear synchrotron and EC cooling, and of non-linear, time-dependent SSC cooling (Schlickeiser, 2009), which is given by
\begin{align}
|\dot{\gamma}(\gamma,t)|_{tot} =& |\dot{\gamma}(\gamma)|_{syn} + |\dot{\gamma}(\gamma)|_{ec} + |\dot{\gamma}(\gamma,t)|_{ssc} \nonumber \\
=& D_0\left( 1+l_{ec} \right) \gamma^2 + A_0\gamma^2 \intl_0^{\infty} \gamma^2 n(\gamma,t) \td{\gamma} \label{eq:appcool} \eqd
\end{align}

The steps to solve the kinetic equation are presented in Schlickeiser et al. (2010). The solution becomes for $\alpha\ll 1$
\begin{align}
n(\gamma,t) = q_0 \HF{\gamma_0-\gamma} \DF{\gamma-\frac{\gamma_0}{1+\frac{t}{\tl}}} \label{eq:appn01} \eqc
\end{align}
and for $\alpha\gg 1$
\begin{align}
n(\gamma,t) =& q_0 \HF{t_c-t} \DF{\gamma-\frac{\gamma_0}{\left( 1+3\alpha^2\frac{t}{\tl} \right)^{1/3}}} \label{eq:appn11} \\
n(\gamma,t) =& q_0 \HF{t-t_c} \DF{\gamma-\frac{\gamma_o}{\alpha_g+\frac{t}{\tl}}} \label{eq:appn12} \eqd
\end{align}
Eq. (\ref{eq:appn01}) is a purely linear solution, while for $\alpha\gg 1$ the solution is divided into a non-linear part and a modified linear part, respectively, {divided at time $t=t_c$ as given in Eq. (\ref{eq:tc}).}

In order to calculate the intensities from Eq. (\ref{eq:appI}), we further need the respective emission powers of single electrons $p_i(\epss,\gamma)$. The SSC power (Schlickeiser, 2002) is given by
\begin{align}
p_{s}(\epss,\gamma) =&  R P_0 (m_ec^2)^2 \epss \nonumber \\
&\times \intl_{0}^{\infty} \td{\eps} \intl_0^{\infty} \td{\gamma} \frac{n(\gamma,t)}{\gamma^2} CS\left( \frac{2\eps}{3\eps_0\gamma^2} \right) \sigma_{KN}(\epss,\eps,\gamma) \label{eq:appsp}  \eqd
\end{align}
The parameters are $P_0 = 2\times 10^{24}$erg$^{-1}$s$^{-1}$, and $\eps_0 = 2.3\times 10^{-14} b$. $\eps$ is the normalized soft photon energy, and $CS(x) \approx a_0 x^{-2/3} e^{-x}$, $a_0 = 1.151275$, describes the isotropic synchrotron photon emissivity (Crusius \& Schlickeiser, 1986).

The isotropic EC power (Dermer \& Schlickeiser, 1993) is calculated by
\begin{align}
p_{e}(\epss,\gamma) = c\epss \intl_0^{\infty} \frac{u_{ec}(\eps)}{\eps} \sigma_{KN}(\epss,\eps,\gamma) \td{\eps} \label{eq:appep} \eqd
\end{align}
As stated before, we assume a line like soft external photon source with normalized line energy $\epec$. In this case the energy density in external photons becomes
\begin{align}
u_{ec}(\eps) = \frac{4\Gamma_b^2}{3} u_{ec}^{\prime} \DF{\eps-\epec} \label{eq:appuec} \eqd
\end{align}

In both SSC and EC emission power we use the full Klein-Nishina cross-section (Blumenthal \& Gould, 1970)
\begin{align}
\sigma_{KN}(\epss,\eps,\gamma) = \frac{3\sigma_T}{4\eps\gamma^2} G(q) \HF{\gamma-\gamma_{min}} \label{eq:appskn} \eqc
\end{align}
with the Thomson cross-section $\sigma_T = 6.65\times 10^{-25}$cm$^{2}$ and
\begin{align}
G(q) =& G_0(q) + \frac{\Gamma_{KN}^2q^2(1-q)}{2(1+\Gamma_{KN}q)} \nonumber \\
=& G_0(q) + 2\eps\epss q(1-q) \label{eq:appG} \eqc \\
G_0(q) =& 2q \ln{q} + (1+2q)(1-q) \label{eq:appG0} \eqc \\
\Gamma_{KN} =& 4\eps\gamma \label{eq:appGamkn} \eqc \\
q =& \frac{\epss}{\Gamma_{KN}(\gamma-\epss)} \label{eq:appq} \eqc \\
\gamma_{min} =& \frac{\epss}{2} \left[ 1+\sqrt{1+\frac{1}{\eps\epss}} \right] \label{eq:appgammin} \eqd
\end{align}

Inserting all these definitions in Eq. (\ref{eq:appI}) yields the intensities given in section \ref{sec:int}.
%
%
\section{Intermediate times approximations}
\subsection{SSC with $\alpha\ll 1$} \label{app:ssca01}
The intermediate time section is here for $\tss{02}<t<\lam_0$. The integral then becomes
\begin{align}
\Ls{\ks,t} =& 6 I_{s,0} \ks^{1/3} \intl_0^{t/\lam0} \left( 1+\frac{t-\lam_0\lam}{t_{lin}} \right)^{4/3} e^{-\ks \left( 1+\frac{t-\lam_0\lam}{t_{lin}} \right)^{4}} \nonumber \\
& \times \left( \lam-\lam^2 \right) \td{\lam} \nonumber \\
=& 6 I_{s,0} \ks^{1/3} \left( 1+\frac{t}{t_{lin}} \right)^{4/3} e^{-\ks \left( 1+\frac{t}{t_{lin}} \right)^{4}} F(\ks,t) \eqc \label{eq:appLs012a}
\end{align}
with the integral
\begin{align}
F(\ks,t) =& \intl_0^{t/\lam_0} \left( 1-\frac{\lam_0\lam}{\tl+t} \right)^{4/3} e^{-\ks\left( 1+\frac{t-\lam_0\lam}{\tl} \right)^4+\ks\left( 1+\frac{t}{\tl} \right)^4} \nonumber \\
& \times \left( \lam-\lam^2 \right) \td{\lam} \eqd
\end{align}
Approximating to first order yields
\begin{align}
F(\ks,t) \approx & \intl_0^{t/\lam_0} \left( 1-\frac{4}{3}\frac{\lam_0\lam}{\tl+t} \right) e^{\frac{4\ks\lam_0}{\tl}\left( 1+\frac{t}{\tl} \right)^3 \lam} \left( \lam-\lam^2 \right) \td{\lam} \nonumber \\
\approx & \frac{\tl}{4\ks\lam_0\left( 1+\frac{t}{\tl} \right)^3} \left( \frac{t}{\lam_0} \right) \left[ 1-\left( 1+\frac{4\lam_0}{3(t+\tl)} \right)\frac{t}{\lam_0} \right] \nonumber \\
& \times e^{\frac{4\ks t}{\tl}\left( 1+\frac{t}{\tl} \right)^3} \eqd 
\end{align}
The second approximation was achieved by integrating by parts and approximating again to first order. Inserting this into Eq. (\ref{eq:appLs012a}) the light curve becomes
\begin{align}
\Ls{\ks,t} \approx \frac{3I_{s,0}\tl}{2\lam_0} \ks^{-2/3} e^{-\ks} \left( \frac{t}{\lam_0} \right) \left[ 1-\frac{t}{\lam_0} \right] 
\end{align}
equalling Eq. (\ref{eq:Ls012}). The approximation $t\ll\tl$ is justified, since the terms with $1+\frac{t}{\tl}$ enter the final solution again by the stitching, if necessary. The important result is the break to a linear dependence of the light curve for times $t>\tss{02}$. 
\subsection{SSC with $\alpha\gg 1$ in the early time limit} \label{app:ssca10a}
The integration is similar to the case $\alpha\ll 1$. For times $\tss{12}<t<t_c<\lam_0$ the integral becomes
\begin{align}
\Ls{\ks,t} =& 6 I_{s,0} \ks^{1/3} \intl_0^{t/\lam0} \left( 1+3\alpha^2\frac{t-\lam_0\lam}{t_{lin}} \right)^{4/9} \nonumber \\
& \times e^{-\ks \left( 1+3\alpha^2\frac{t-\lam_0\lam}{t_{lin}} \right)^{4/3}} \left( \lam-\lam^2 \right) \td{\lam} \nonumber \\
=& 6 I_{s,0} \ks^{1/3} \left( 1+3\alpha^2\frac{t}{t_{lin}} \right)^{4/9} e^{-\ks \left( 1+3\alpha^2\frac{t}{t_{lin}} \right)^{4/3}} \nonumber \\
&\times F(\ks,t) \eqd \label{eq:appLs112a}
\end{align}
Here the integral function $F(\ks,t)$ can be approximated as
\begin{align}
F(\ks,t) =& \intl_0^{t/\lam_0} \left( 1-\frac{\lam_0\lam}{\tla+t} \right)^{4/9} e^{-\ks\left( 1+3\alpha^2\frac{t-\lam_0\lam}{\tl} \right)^{4/3}} \nonumber \\
& \times e^{\ks\left( 1+3\alpha^2\frac{t}{\tl} \right)^{4/3}} \left( \lam-\lam^2 \right) \td{\lam} \nonumber \\
\approx & \intl_0^{t/\lam_0} \left( 1-\frac{4}{9}\frac{\lam_0\lam}{\tla+t} \right) e^{\frac{4\alpha^2\ks\lam_0}{\tl} \lam} \left( \lam-\lam^2 \right) \td{\lam} \nonumber \\
\approx & \frac{\tl}{4\alpha^2\ks\lam_0} \left( \frac{t}{\lam_0} \right) \left[ 1-\left( 1+\frac{4\lam_0}{9(t+\tla)} \right)\frac{t}{\lam_0} \right] \nonumber \\
& \times e^{\frac{4\alpha^2\ks t}{\tl}} \eqd 
\end{align}
Inserting yields
\begin{align}
\Ls{\ks,t} \approx \frac{3I_{s,0}\tl}{2\alpha^2\lam_0} \ks^{-2/3} e^{-\ks} \left( \frac{t}{\lam_0} \right) \left[ 1-\frac{t}{\lam_0} \right] \eqc
\end{align}
which is the same as Eq. (\ref{eq:LsIA1aii}). As before, the exact form of {this solution is} not important, because of the stitching.
\subsection{SSC with $\alpha\gg 1$ in the late time limit} \label{app:ssca10b}
As before, the integral for times $\tss{22}<t<t_c+\lam_0$ results in
\begin{align}
\Ls{\ks,t} =& 6 I_{s,0} \ks^{1/3} \intl_0^{\frac{t-t_c}{\lam0}} \left( \alpha_g+\frac{t-\lam_0\lam}{t_{lin}} \right)^{4/3} \nonumber \\
& \times e^{-\ks \left( \alpha_g+\frac{t-\lam_0\lam}{t_{lin}} \right)^{4}} \left( \lam-\lam^2 \right) \td{\lam} \nonumber \\
=& 6 I_{s,0} \ks^{1/3} \left( \alpha_g+\frac{t}{t_{lin}} \right)^{4/3} e^{-\ks \left( \alpha_g+\frac{t}{t_{lin}} \right)^{4}} F(\ks,t) \eqd \label{eq:appLs211a}
\end{align}
Here the integral function $F(\ks,t)$ can be approximated as
\begin{align}
F(\ks,t) =& \intl_0^{\frac{t-t_c}{\lam_0}} \left( 1-\frac{\lam_0\lam}{\tl\alpha_g+t} \right)^{4/3} e^{-\ks\left( \alpha_g+\frac{t-\lam_0\lam}{\tl} \right)^{4}} \nonumber \\
& \times e^{\ks\left( \alpha_g+\frac{t}{\tl} \right)^{4}} \left( \lam-\lam^2 \right) \td{\lam} \nonumber \\
\approx & \intl_0^{\frac{t-t_c}{\lam_0}} \left( 1-\frac{4}{3}\frac{\lam_0\lam}{\tl\alpha_g+t} \right) e^{\frac{4\ks\lam_0}{\tl}\left( \alpha_g+\frac{t}{\tl} \right)^3 \lam} \nonumber \\
&\times \left( \lam-\lam^2 \right) \td{\lam} \nonumber \\
\approx & \frac{\tl}{4\ks\lam_0 \left( \alpha_g+\frac{t}{\tl} \right)^3} \left( \frac{t-t_c}{\lam_0} \right) \nonumber \\
& \times \left[ 1-\left( 1+\frac{4\lam_0}{3(t+\tl\alpha_g)} \right)\frac{t-t_c}{\lam_0} \right] e^{\frac{4\ks (t-t_c)}{\tl}} \eqd 
\end{align}
Inserting yields
\begin{align}
\Ls{\ks,t} \approx \frac{3I_{s,0}\tl}{2\alpha^{5/3}\lam_0} \ks^{-2/3} e^{-\alpha^4\ks} \left( \frac{t-t_c}{\lam_0} \right) \left[ 1-\frac{t}{t_c+\lam_0} \right] \eqc
\end{align}
which is the same as Eq. (\ref{eq:LsII1aiAa}). The last approximation used $t\rightarrow t_c$. Noting as before, the exact form of this result is not important, apart from the linear time dependence.
\subsection{EC with $\alpha\ll 1$} \label{app:eca01}
The integral of the intermediate time domain $\tsu{00}<t<\lam_0$ for energies $\epss>\epsu{00}$ is given by
\begin{align}
\Le =& 6 I_{e,0} \epss \intl_0^{t/\lam_0} \left( 1+\frac{t-\lam_0\lam}{\tl} \right)^2 G\left( q\left( \frac{t-\lam_0\lam}{\tl} \right) \right) \nonumber \\
&\times  \left( \lam-\lam^2 \right) \td{\lam} \eqd
\end{align}
Using $q$ instead of $\lam$ as the integration variable, the substitution becomes
\begin{align}
\lam(q) = \frac{t+\tl}{\lam_0} - \frac{\Gamec\tl}{2\lam_0}q \left( \sqrteeq{} -1  \right) \eqd
\end{align}
Hence, the new lower limit $A = \left[ \Gamec\left( \frac{\gamma_0}{\epss} -1 \right) \right]^{-1}$, and the integral
\begin{align}
\Le =& \frac{3}{8 \lam_0} I_{e,0} \tl \Gamec^3 \epss \intl_{A}^{1} q^2 \frac{\left( \sqrteeq{} -1 \right)^4}{\sqrteeq{}} \nonumber \\
&\times G(q) \lam(q) (1-\lam(q)) \td{q} \nonumber \\
=& \frac{3}{8} I_{e,0} \Gamec^3 \epss \frac{\tl}{\lam_0} \left( \frac{t}{\lam_0} \right) \intl_{A}^{1} q^2 \frac{\left( \sqrteeq{} -1 \right)^4}{\sqrteeq{}} \nonumber \\
&\times G(q) \lam_{e,0}^{*}(t) \left( 1-\frac{t}{\lam_0} \lam_{e,0}^{*}(t) \right) \td{q} \nonumber \\
\approx & \frac{3}{8} I_{e,0} \Gamec^3 \epss \frac{\tl}{\lam_0} \left( \frac{t}{\lam_0} \right) \left[ 1-\frac{t}{\lam_0} \right] \nonumber \\
&\times \intl_{A}^{1} q^2 \frac{\left( \sqrteeq{} -1 \right)^4}{\sqrteeq{}} G(q) \td{q} \nonumber \\
=& \frac{3}{8} I_{e,0} \Gamec^3 \epss F_0(\epss) \frac{\tl}{\lam_0} \left( \frac{t}{\lam_0} \right) \left[ 1-\frac{t}{\lam_0} \right] \eqd
\end{align}
The approximation $t\gg\tl$ is performed for the function
\begin{align}
\lam_{e,0}^{*}(t) =& 1+ \frac{\tl}{t} - \frac{\Gamec\tl}{2t}q \left( \sqrteeq{} -1  \right) \nonumber \\
\approx & 1 \eqc
\end{align}
which is a valid assumption here, since for a reasonable large energy domain $\epsu{00}<\epss<\Gamec\gamma_0/(1+\Gamec)$ we find $\lam_0>\tl$. The integral $F_0(\epss)$ could be solved with the methods of Zacharias \& Schlickeiser (2012b). However, we are only interested in the time-dependency here, and therefore the complete derivation of the integral is useless for our purpose. Thus, we have obtained the light curve given in Eq. (\ref{eq:Le014}).
\subsection{EC with $\alpha\gg 1$ in the early time limit} \label{app:eca10a}
The intermediate time domain integral for the EC light curve for $\alpha\gg 1$ in the early time limit becomes
\begin{align}
\Le =& 6 I_{e,0} \epss \intl_{\frac{t-t_c}{\lam_0}}^{t/\lam_0} \left( 1+3\alpha^2\frac{t-\lam_0\lam}{\tl} \right)^{2/3} \nonumber \\
&\times G\left( q_1\left( 3\alpha^2\frac{t-\lam_0\lam}{\tl} \right) \right)  \left( \lam-\lam^2 \right) \td{\lam} \eqd
\end{align}
Substituting $q_1$ for $\lam$ yields
\begin{align}
\lam(q_1) =& \frac{t}{\lam_0} + \frac{\tl}{3\alpha^2\lam_0} \nonumber \\
& - \frac{\tl}{3\alpha^2\lam_0} \left( \frac{\Gamec}{2}q_1 \right)^3 \left( \sqrteeq{1} -1  \right)^3 \eqd
\end{align}
Thus, the new lower limit $A = \epss / \left[ \Gamec\left( \gamma_0-\epss \right) \right]$, the new upper limit $B = \epss\alpha^2 / \left[ \Gamec\left( \gamma_0-\alpha\epss \right) \right]$, and the light curve
\begin{align}
\Le =& \frac{3 I_{e,0} \Gamec^5}{32\alpha^2} \epss \frac{\tl}{\lam_0} \left( \frac{t}{\lam_0} \right) \intl_{A}^{B} q_1^4 \frac{\left( \sqrteeq{1} -1 \right)^6}{\sqrteeq{1}} \nonumber \\
&\times G(q_1) \lam_{e,1}^{*}(t) \left( 1-\frac{t}{\lam_0} \lam_{e,1}^{*}(t) \right) \td{q_1} \nonumber \\
\approx & \frac{3 I_{e,0} \Gamec^5}{32\alpha^2} \epss \frac{\tl}{\lam_0} \left( \frac{t}{\lam_0} \right) \left[ 1-\frac{t}{\lam_0} \right] \nonumber \\
&\times \intl_{A}^{B} q_1^4 \frac{\left( \sqrteeq{1} -1 \right)^6}{\sqrteeq{1}} G(q_1) \td{q_1} \nonumber \\
=& \frac{3 I_{e,0} \Gamec^5}{32\alpha^2} \epss F_1(\epss) \frac{\tl}{\lam_0} \left( \frac{t}{\lam_0} \right) \left[ 1-\frac{t}{\lam_0} \right] \eqd
\end{align}
The function
\begin{align}
\lam_{e,1}^{*}(t) =& 1+ \frac{\tl}{3\alpha^2 t} - \frac{\tl}{3\alpha^2 t} \left( \frac{\Gamec}{2} q_1 \right)^3 \left( \sqrteeq{1} -1  \right)^3 \nonumber \\
\approx & 1 
\end{align}
is again approximated for small $t$.

As in the case $\alpha\ll 1$ one obtains a linear time dependence for the light curve, which also contains an integral $F_1$ solely depending on $\epss$. As discussed above, we focus on the time-dependency. For plotting purposes the function $F_1(\epss)$ could be evaluated  with the methods of Zacharias \& Schlickeiser (2012b) or numerically.
\subsection{EC with $\alpha\gg 1$ in the late time limit} \label{app:eca10b}
In the intermediate time domain $\tsu{20}<t<t_c+\lam_0$ the EC light curve in the late time limit for $\alpha\gg 1$ equals
\begin{align}
\Le =& 6 I_{e,0} \epss \intl_{\frac{t-\tsu{20}}{\lam_0}}^{\frac{t-t_c}{\lam_0}} \left( \alpha_g+\frac{t-\lam_0\lam}{\tl} \right)^2 \nonumber \\
&\times G\left( q_2\left( \frac{t-\lam_0\lam}{\tl} \right) \right) \left( \lam-\lam^2 \right) \td{\lam} \eqd
\end{align}
Using $q_2$ instead of $\lam$ as the integration variable, the substitution is
\begin{align}
\lam(q_2) = \frac{t}{\lam_0} + \frac{\tl\alpha_g}{\lam_0} - \frac{\Gamec\tl}{2\lam_0}q_2 \left( \sqrteeq{2} -1  \right) \eqd
\end{align}
The lower limit becomes $A = \epss\alpha^2 / \left[ \Gamec\left( \gamma_0 - \epss\alpha \right) \right]$ giving
\begin{align}
\Le =& \frac{3\tl}{8 \lam_0} I_{e,0} \Gamec^3 \epss \intl_{A}^{1} q_2^2 \frac{\left( \sqrteeq{2} -1 \right)^4}{\sqrteeq{2}} \nonumber \\
&\times G(q_2) \lam_{e,2}^{*}(t) \left( 1-\frac{t}{\lam_0} \lam_{e,2}^{*}(t) \right) \td{q_2} \nonumber \\
\approx & \frac{3}{8} I_{e,0} \Gamec^3 \epss \frac{\tl}{\lam_0} \left( \frac{t}{\lam_0} \right) \left[ 1-\frac{t}{\lam_0+t_c} \right] \nonumber \\
&\times \intl_{A}^{1} q_2^2 \frac{\left( \sqrteeq{2} -1 \right)^4}{\sqrteeq{2}} G(q_2) \td{q_2} \nonumber \\
=& \frac{3}{8} I_{e,0} \Gamec^3 \epss F(\epss) \frac{\tl}{\lam_0} \left( \frac{t}{\lam_0} \right) \left[ 1-\frac{t}{\lam_0} \right] \eqd
\end{align}
Exploiting again that
\begin{align}
\lam_{e,2}^{*}(t) =& 1+ \frac{\tl\alpha_g}{t} - \frac{\Gamec\tl}{2t}q_2 \left( \sqrteeq{2} -1  \right) \nonumber \\
\approx & 1
\end{align}
for small $t$, the linear time-dependence of the intermediate time regime is recovered, again. The discussion concerning $F_2(\epss)$ does not need to be repeated here.
%
%

%
%
%
\end{document}